\documentclass{aa}  

\usepackage{graphicx}
\usepackage{mathabx}
\usepackage{textgreek}
\usepackage{soul}
\usepackage{ulem}
\newcommand{\obli}{\textlambda~}
\usepackage{txfonts}
\usepackage[breaklinks=true]{hyperref}
\hypersetup{
    colorlinks=true,
    linkcolor=blue,
    filecolor=magenta,      
    urlcolor=magenta,
    citecolor=blue,
    pdftitle={HD332231b},
    pdfpagemode=FullScreen,
    }

\begin{document}

   \title{Moderately misaligned orbit of the warm sub-Saturn HD\,332231\,b}


   \author{E. Sedaghati
          \inst{1,2,3}
          \and
          A. S\'anchez-L\'opez
          \inst{4}
          \and
          S. Czesla
          \inst{5}
          \and
          M. L\'opez-Puertas
          \inst{1}
          \and
          P.\,J. Amado
          \inst{1}
          \and
          E. Palle
          \inst{6,7}
          \and\\
          K. Molaverdikhani
          \inst{8,9,10}
          \and
          J.\,A. Caballero
          \inst{11}
          \and
          L. Nortmann
          \inst{12}
          \and
          A. Quirrenbach
          \inst{10}
          \and
          A. Reiners
          \inst{12}
          \and
          I. Ribas
          \inst{13,14}
          }

   \institute{Instituto de Astrof\'isica de Andaluc\'ia (IAA-CSIC), Glorieta de la Astronom\'ia s/n, 18008 Granada, Spain
        \and
            Facultad de Ingenier\'ia y Ciencias, Universidad Adolfo Ib\'a\~nez, Av.\ Diagonal las Torres 2640, Pe\~nalol\'en, Santiago, Chile
        \and
            Millennium Institute for Astrophysics, Chile; \email{esedagha@astrofisica.cl}
        \and
            Leiden Observatory, Leiden University, Postbus 9513, 2300 RA, Leiden, The Netherlands
        \and
             Hamburger Sternwarte, Universit\"at Hamburg, Gojenbergsweg 112, 21029 Hamburg, Germany
        \and
            Instituto de Astrofísica de Canarias (IAC), 38200 La Laguna, Tenerife, Spain
        \and
            Deptartamento de Astrofísica, Universidad de La Laguna (ULL), 38206 La Laguna, Tenerife, Spain
        \and
            Universitäts-Sternwarte, Ludwig-Maximilians-Universität München, Scheinerstrasse 1, 81679 München, Germany
        \and
            Exzellenzcluster Origins, Boltzmannstraße 2, 85748 Garching, Germany
        \and
            Landessternwarte, Zentrum für Astronomie der Universität Heidelberg, Königstuhl 12, 69117 Heidelberg, Germany
        \and
            Centro de Astrobiología (CSIC-INTA), ESAC, Camino bajo del castillo s/n, 28692 Villanueva de la Cañada, Madrid, Spain
        \and
            Institut für Astrophysik, Georg-August-Universität, FriedrichHund-Platz 1, 37077 Göttingen, Germany
        \and
            Institut de Ciències de l’Espai (ICE, CSIC), Campus UAB, C/Can Magrans s/n, 08193 Bellaterra, Spain
        \and
            Institut d’Estudis Espacials de Catalunya (IEEC), 08034 Barcelona, Spain
             }

   \date{Received 18 October 2021 / Accepted 17 November 2021}

 
  \abstract
    {Measurements of exoplanetary orbital obliquity angles for different classes of planets are an essential tool in testing various planet formation theories. Measurements for those transiting planets on relatively large orbital periods ($P$\,$>$\,10\,d) present a rather difficult observational challenge. Here we present the obliquity measurement for the warm sub-Saturn planet HD\,332231\,b, which was discovered through Transiting Exoplanet Survey Satellite (TESS) photometry of sectors 14 and 15, on a relatively large orbital period (18.7\,d). Through a joint analysis of previously obtained spectroscopic data and our newly obtained CARMENES transit observations, we estimated the spin-orbit misalignment angle, \textlambda, to be $-42.0^{+11.3}_{-10.6}$\,$\deg$, which challenges Laplacian ideals of planet formation. Through the addition of these new radial velocity (RV) data points obtained with CARMENES, we also derived marginal improvements on other orbital and bulk parameters for the planet, as compared to previously published values. We showed the robustness of the obliquity measurement through model comparison with an aligned orbit. Finally, we demonstrated the inability of the obtained data to probe any possible extended atmosphere of the planet, due to a lack of precision, and place the atmosphere in the context of a parameter detection space.}


   \maketitle
%

\section{Introduction}
\label{sec: Introduction}
Relative to the invariable plane of Laplace \citep{Laplace1796}, which is orthogonal to the angular momentum vector and passes through the barycentre, both terrestrial planets and gas giants in our Solar System exist within well-aligned orbits. This obliquity angle is within $\sim$\,2\,$\deg$ for all planets, with the exception of Mercury \citep[\textlambda$_{\Mercury} \approx 6.3$\,$\deg$;][]{Souami2012}. This picture, however, is drastically different for the minor bodies in the Solar System. This coplanarity observed in our own backyard paved the way for understanding the formation and evolution of planetary systems, which is tied to the acceptance of the heliocentric model \citep{kuiper1951,gingerich1973}.

\begin{figure*}
    \centering
    \includegraphics[width=\linewidth]{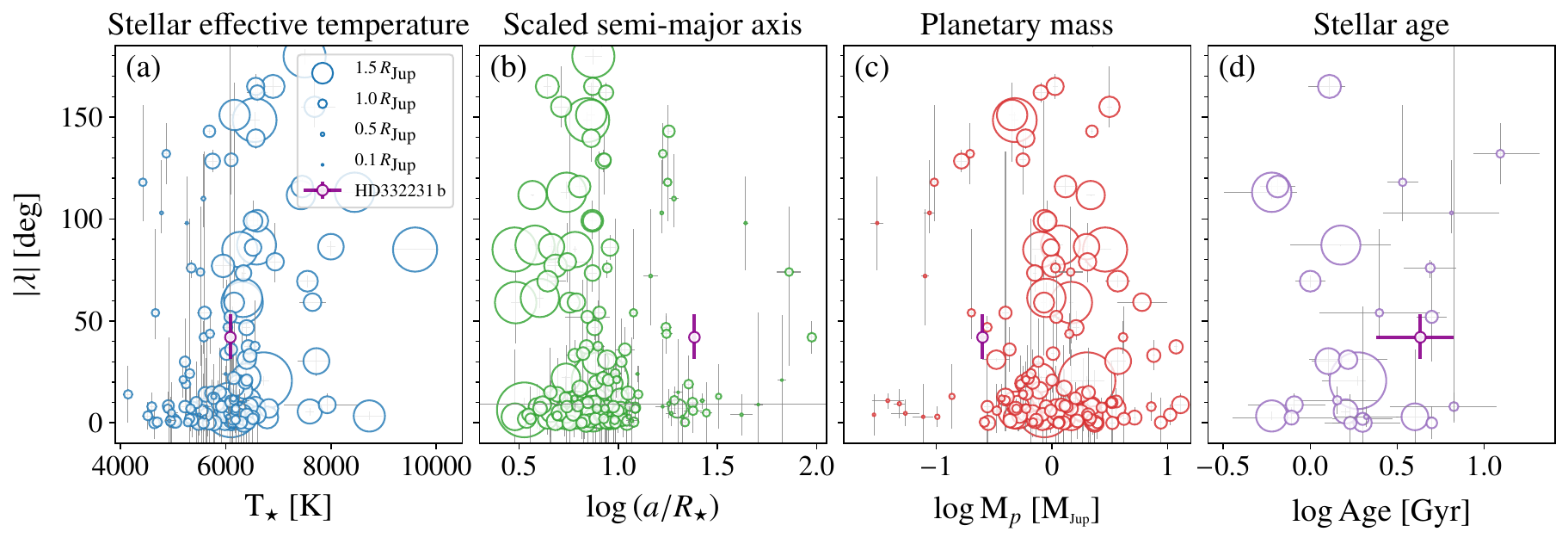}
    \caption{Orbital obliquity distributions of those transiting exoplanets with values determined from the Rossiter-McLaughlin anomaly. (a) shows the distribution of \obli with the temperature of the host star, where it is quite evident that giant planets around hot stars are almost entirely on misaligned orbits. (b) represents the distribution of \obli with the scaled semi-major axis, whereby giant planets whose orbital semi-major axes are $\lesssim 10\,R_\star$ tend to be on highly misaligned orbits. (c) shows again the same distribution, but for the mass of the planet where the more massive planets present a larger sample of misaligned orbits. (d) represents the distribution of \obli as a function of the estimated stellar age for those targets where an estimation is available (from NASA exoplanet archive September 2021). In all panels HD\,332231\,b is represented as the purple point.}
    \label{fig:Obliquity_dist}
\end{figure*}

Discovery of exoplanetary systems has presented a rather more complex picture of planetary architectures. Transiting exoplanets, those that cross the visible disk of their host stars from our vantage point, permit the measurement of the spin-orbit misalignment between the planetary orbital plane and the stellar equatorial plan that is projected onto the sky, as well as other crucial characteristics \citep{Triaud2018}. This is what is referred to as the sky-projected obliquity angle (\obli hereafter). Its measurement is performed through the observations of the Rossiter-McLaughlin effect \citep{Rossiter1924,McLaughlin1924}. For exoplanets, it entails observations of the stellar radial velocity (RV) during the planetary transit. The anomaly in the measured RV values arises from the deformation of absorption lines from which they are determined, which is caused by the transiting planet occulting either the blue- or red-shifted portion of the spinning stellar disk. The measurement of this effect is possible thanks to precision, high dispersion spectrographs at large telescopes that allow for one to obtain high resolution and large signal-to-noise (S/N) spectra at relatively high temporal sampling. The measurement of \obli has been performed for a large number of transiting exoplanets, full details of which can be found in TEPCAT\footnote{\url{www.astro.keele.ac.uk/jkt/tepcat/obliquity.html}} \citep{Southworth2011}. These have revealed a surprising diversity in the orbital alignments \citep[for example][]{Queloz2000,Winn2005,Winn2006,Winn2009,Triaud2009,Gandolfi2010,Mancini2018,Yu2018,Lendl2020,Sedaghati2021}, which is in contrast to the Laplacian ideals of planets forming inside a flat disk, coplanar with the stellar equator and staying there \citep{Laplace1796}. A surprising picture that has emerged is that a significant fraction of those close-in hot-Jupiter regime planets are on misaligned orbits, as is evident in panel (b) of Fig. \ref{fig:Obliquity_dist} \citep{Albrecht2012,Dawson2014}. Furthermore, the spectral type of the host also appears to play a role, whereby giant planets around hot stars seem to exist on more oblique orbits (panel (a) of Fig.\,\ref{fig:Obliquity_dist}), perhaps pointing to a different, more chaotic formation history, as compared to their cooler counterparts. The relation between the obliquity and the host star temperature was observed by \citet{Winn2010}, who placed the boundary between the two regimes at T$_\star$\,=\,6250\,K \citep[namely the Kraft break;][]{Kraft1967}. \citet{Hebrard2011} also point out a lack of planets with masses >\,3M$_{\textrm{Jup}}$ on retrograde orbits, the distribution for which is shown in panel (c) of Fig.\,\ref{fig:Obliquity_dist}. Tidal interactions over time with the host star are also expected to realign orbits of close-in, massive planets \citep{Zahn1977}. Attempts have been made to study the impact of stellar age on the obliquity of planetary orbits \citep[for example][]{Safsten2020,Anderson2021}, with \citet{Triaud2011} finding that hot-Jupiters around younger A stars are more misaligned, setting the age barrier at 2.5\,Gyr. However, a lack of precision in the measured stellar ages and the absence of uniform and homogeneous studies estimating those ages have hindered any concrete conclusions being drawn with regard to the impact of stellar ages on planetary orbital alignments. This fact is evident in panel (d) of Fig. \ref{fig:Obliquity_dist}.

The underlying causes of the aforementioned misalignment for the hot-Jupiter sample are a subject of debate. One hypothesis suggests that during the high eccentricity migration, through which hot Jupiters are formed \citep{Petrovich2015,Dawson2018}, dynamical interactions contribute to increasing the obliquity of planetary orbit, pushing it away from the initial equatorial plane. Such dynamical interactions are attributed to phenomena such as secular planet-planet interactions \citep{Naoz2011} due to Kozai-Lidov cycles \citep{Katz2011,Storch2014}, planet-planet scattering \citep{Rasio1996,Marzari2014} or secular chaos \citep{Laskar1993,Millholland2019}. There exists, however, another school of thought that does not invoke any mechanisms related to planetary migration. For instance, torques from wide-orbiting companions \citep{Batygin2012,Huber2013}, or a chaotic star formation environment \citep{Bate2010}, among others, have been suggested as responsible for the observed misalignment of the hot Jupiter class planets. 

Giant planets on wider orbits, those typically with orbital periods ranging from 10 to 100 days, are classified as warm. Warm Jupiters and Saturns are typically subject to much weaker tidal interactions as compared to their hot counterparts. \citet{Dong2014} suggest that planets in this regime form through high eccentricity migration, overcoming the precession caused by general relativistic effects. Another alternative, of course, is the Laplacian framework of these planets forming quiescently within aligned disks, and in all likelihood a combination of these two theories is responsible for the observed sample. Measurements of obliquity angles for a large sample of giant planets in the warm regime can somewhat help differentiate between the two competing schools of thought presented above. Namely, according to \citet{Dong2014} if the misalignments are due to high eccentricity migration alone, then the majority of warm giant planets are expected to be on aligned orbits, as predicted by the Laplace mechanism. This statement is particularly true for low eccentricity warm-Jupiters \citep[$e$\,$\lesssim$\,$0.2$;][]{Dong2014} and would favour arguments invoking migration mechanisms. However, if the same pattern of misalignment is observed for both classes, then explanations not involving planetary migration are favoured. Once a large sample of \obli values for the warm class is obtained, the interpretation of its distribution might be simpler, as the orbital architectures are expected to be more pristine. This is due to the fact that for the close-in hot Jupiters this information is most likely lost because of significant dynamical interactions with the host star.

From an observational point of view, measurement of \obli for warm giant planets presents a more difficult challenge as compared to the hot class. This is due to the fact that they are less probable to transit, offer relatively fewer observable transits in a given period of time and their long transit durations make the observations of a complete transit from the ground rather cumbersome. It is noted that Rossiter-McLaughlin observations, for now, are only feasible from ground-based observatories as they require very high resolution, precision spectroscopy.

In this study we present the obliquity measurement of the warm sub-Saturn planet, HD\,332231\,b \citep{Dalba2020}, discovered from NASA's Transiting Exoplanet Survey Satellite \citep[TESS; ][]{Ricker2015} photometry of sectors 14 and 15. It has a mass and radius of 0.244\,$\pm$\,0.021\,M$_\textrm{Jup}$ and 0.867\,$\pm$\,0.026\,\,R$_\textrm{Jup}$ respectively, and orbits a main-sequence F8, 8.56 m$_\textrm{V}$ star on an 18.71\,d ($\pm$\,1.1\,min) circular orbit. 

In what follows we present the observations and data preparation, including telluric correction of the spectra, in section \ref{sec: Observation}, calculation of the RV values, modelling the Rossiter-McLaughlin effect to determine the obliquity of the orbit, as well as transmission spectroscopy to search for atmospheric signals in section \ref{sec: Analysis}, discuss possible implications in section \ref{sec: Discussion}, and finally conclude the study in section \ref{sec: Conclusions}. 

\section{Observations and data preparation}
\label{sec: Observation}

\begin{figure}
    \centering
    \includegraphics[width=\linewidth]{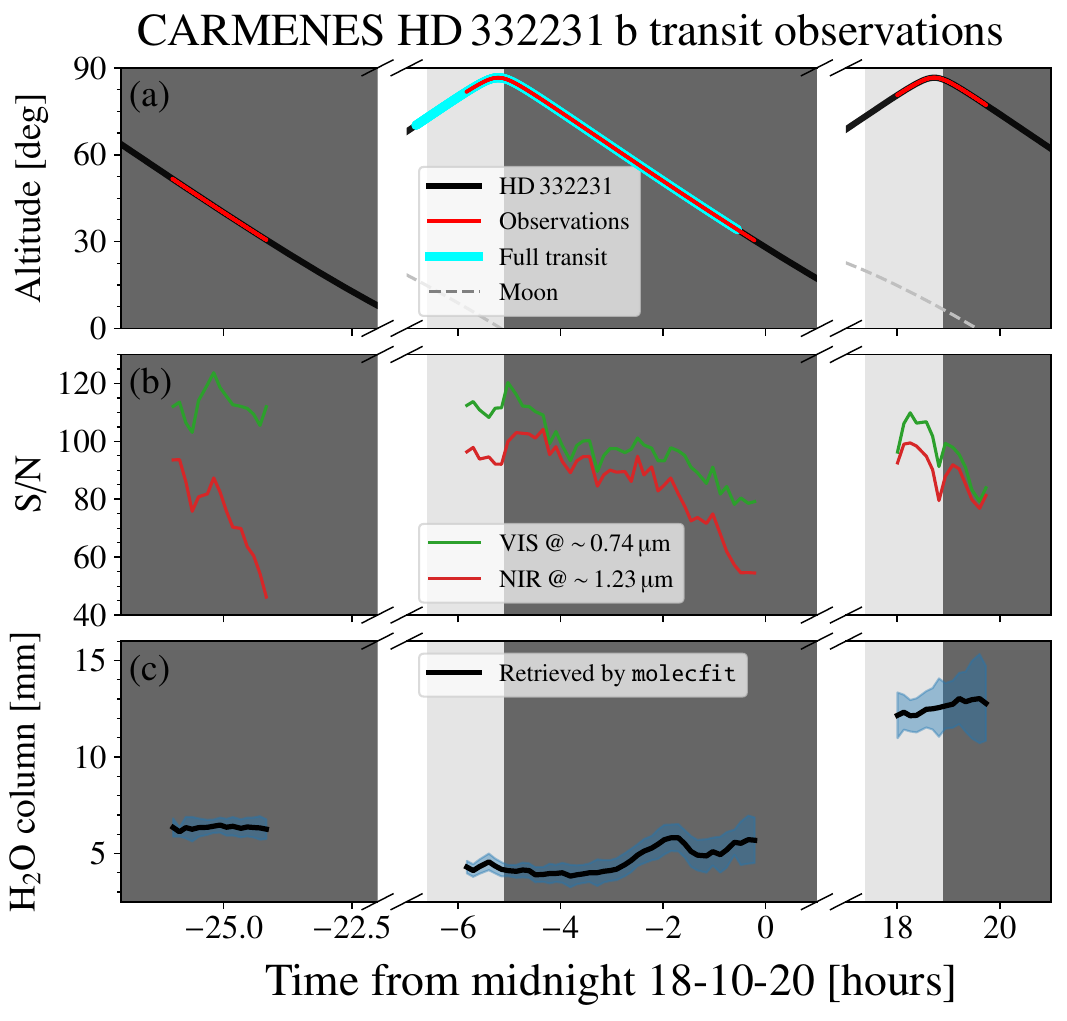}
    \caption{Geometry and conditions for CARMENES observations of HD\,332231\,b transit. (a) presents the altitude of the target at the observatory, whereby the red lines present the duration of the observations, with the cyan line representing the full exoplanetary transit. (b) shows the S/N of the spectra calculated by the pipeline in the VIS and NIR channels, at the given specific wavelengths. (c) shows the retrieved water vapour column density by {\tt molecfit} in mm, with the 1\textsigma~confidence shaded in light blue. In all panels the dark and light shaded regions are the astronomical night and twilight, respectively, whereby the time axis has been broken for better presentation and midnight is in UT. Please note that no atmospheric seeing values were recorded at the observatory for the observations and the cause of the discrepancy in the S/N values of the first night observations in the VIS and NIR channels is not entirely clear.}
    \label{fig: Observations}
\end{figure}

We observed a single transit of HD\,332231\,b on the night of 18-10-2020 spanning the entire night and consequently out of transit observations were obtained during the two nights on either side. The observations were performed with the high dispersion \'echelle spectrograph CARMENES \citep{Quirrenbach2014} installed at the 3.5m telescope at the Calar Alto Observatory in southern Spain. The instrument consists of two spectrographs covering the visible (VIS) and the near-infrared (NIR) channels separately. The VIS channel covers the range of 0.52 to 0.96 \textmu m, with resolving power $\mathcal{R}$\,$\sim$\,94\,600, encompassing 55 spectral orders. The NIR channel covers 0.96 to 1.71 \textmu m within 28 orders at $\mathcal{R}$\,$\sim$\,80\,400. The instrument offers two fibres for light injection, where fibre A is placed on the astronomical source and fibre B is typically used for calibration purposes. For our observations we placed fibre B on sky to monitor and possibly correct for atmospheric emission lines, as well as remove any potential lunar light contamination. The observations were performed with 400s exposure time in the VIS channel and 406s in the NIR, to account for different detector readout times in the two channels. 43 exposures were taken on the night of 18-10, of which only the last 4 were out of transit, as well as 15 and 14 exposures on the nights before and after, respectively. The on-sky configuration of the observations are shown in panel (a) of Fig. \ref{fig: Observations}.

\subsection{Data reduction}
The recorded spectra were subsequently reduced using the {\tt caracal} reduction pipeline \citep[CARMENES Reduction And Calibration;][]{Caballero2016}, which considers all standard astronomical data reduction steps, including a flat-relative optimal extraction algorithm \citep{Zechmeister2014}. The wavelength calibration is performed using Th-Ne and U-Ne lamps for the VIS and NIR channels, respectively, both in combination with a Fabry-P\'erot etalon. The wavelengths were given in the observatory frame and in vacuum. The median S/N values for the in-transit spectra (the night of 18-10) are 98 ($\sim$\,0.74\,\textmu m) and 90 ($\sim$\,1.23\,\textmu m), for the VIS and NIR channels, respectively. The variations of the S/N values for both channels are presented in panel (b) of Fig. \ref{fig: Observations}.

\subsection{Telluric correction}
\label{sec: Telluric}
Stellar spectra obtained with ground-based facilities are always inherently affected by the presence of the Earth's atmosphere. This telluric impact manifests itself as both absorption and emission features, which significantly depend on the location of the observatory and local meteorological conditions. In particular, the atmospheric humidity and temperature profiles along the line of sight of the observations dictate the specific shapes and depths of those telluric lines, which are highly time-dependent. For our observations the variations of humidity along the line of sight, obtained from the modelling routine presented below, are shown in panel (c) of Fig. \ref{fig: Observations}.

In this study we present results from the initial set of spectra, as well as those that have been corrected for the telluric absorption features. The emission features, most prominent in the NIR channel, are simply masked out. For the correction of the telluric absorption features we used ESO's {\tt molecfit} routines \citep{Kausch2015,Smette2015} version 1.5.9, which synthetically model the lines using a line-by-line radiative transfer model. In order to model line shapes and depths, the code requires the atmospheric pressure profile, the initial guess for which it calculates from the humidity and temperature profiles that it obtains from the GDAS\footnote{\url{https://www.ncei.noaa.gov/products/weather-climate-models/global-data-assimilation}} (Global Data Assimilation System) database. The optimal atmospheric profiles are then obtained through minimisation routines, given the recorded spectra. As {\tt molecfit} requires one-dimensional spectra, we stitched all the spectral orders from a single exposure together, creating one spectrum for each pair of VIS and NIR observation.

\begin{figure*}
    \centering
    \includegraphics[width=\linewidth]{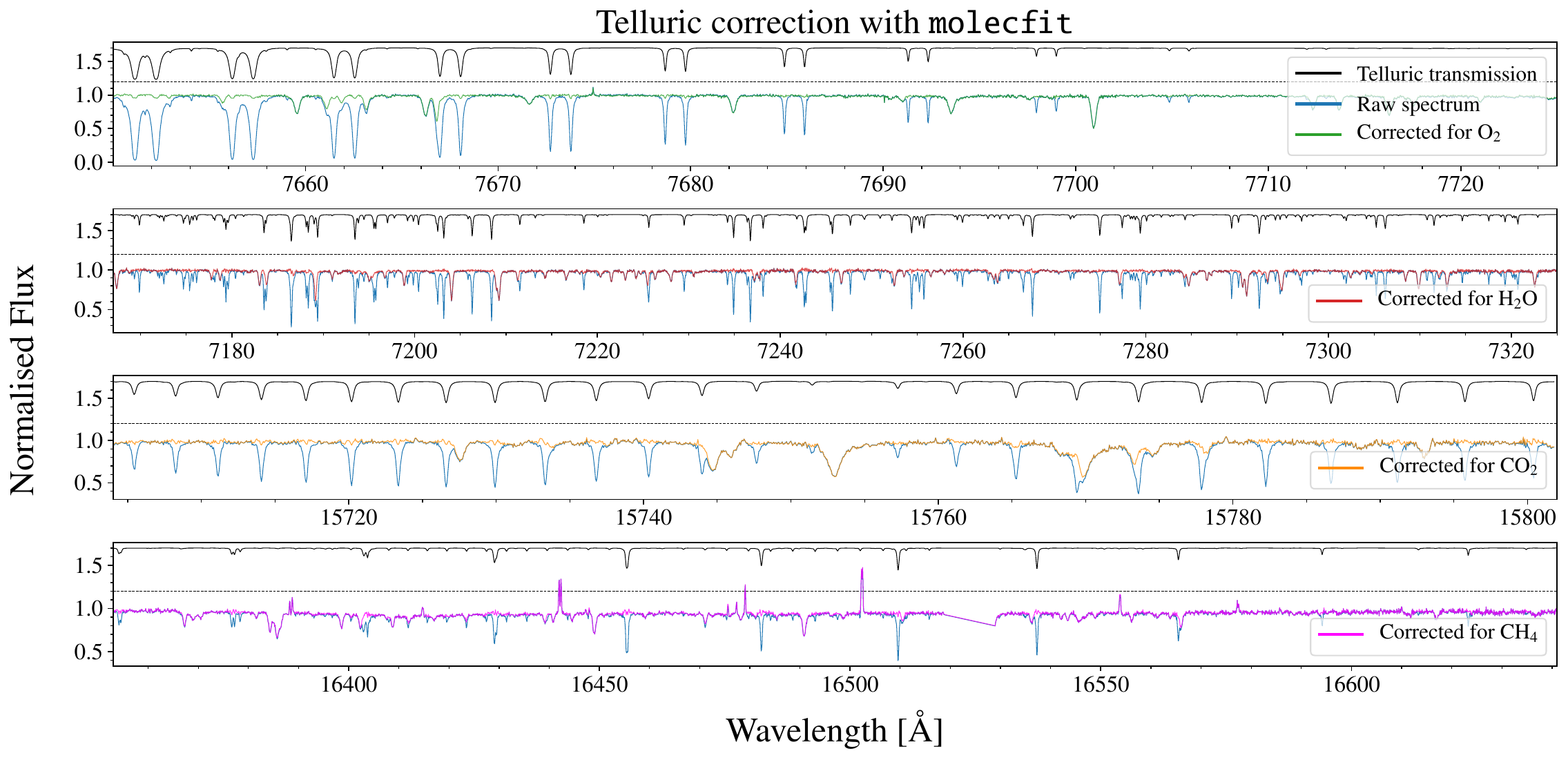}
    \caption{Process of telluric absorption correction of the CARMENES spectra using {\tt molecfit}. The panels show zooms into regions where the main absorbers are, those being O$_2$, H$_2$O, CO$_2$ and CH$_4$, from top to bottom respectively. The modelled telluric transmission spectrum is presented as a solid black line, shifted vertically for clarity, where the zero flux level is indicated with the dashed line. In the bottom panel some sky emission lines are clearly visible, which are masked at the analysis stage.}
    \label{fig: Telluric}
\end{figure*}

For the fitting algorithm of {\tt molecfit}, we included the main atmospheric optical and NIR absorbers of O$_2$, H$_2$O, CO$_2$ and CH$_4$, allowed the continuum to be modelled with a local polynomial of the 4th order and relative convergence criterion of $10^{-10}$. The line shapes were modelled using a Voigt profile, which is assumed as variable to account for the instrumental profile variation with wavelength. Example regions for all the fitted species are shown in Fig. \ref{fig: Telluric}.

In order to remove the telluric absorption signature from the original two-dimensional spectra (\'echelle order-by-order data matrices), the corrected spectrum obtained from {\tt molecfit} was resampled on to the wavelength grids of the individual \'echelle orders and stored as fits files for further analysis.

As an alternative to correcting the telluric absorption with {\tt molecfit}, we also used the {\tt SYSREM} algorithm, which is equivalent to a Principal Component Analysis (PCA) algorithm for unequal uncertainties \citep{Tamuz2005}, to remove the pseudo-stationary telluric and stellar absorption lines from the spectral matrices. This approach has been shown in a number of studies to effectively remove the unwanted stellar and Earth atmospheric signals from the recorded spectra \citep[for example][]{Birkby2013,SanchezLopez2020}. However, we chose not to further explore the {\tt SYSREM} reduced data, neither for the calculation of RVs nor for the atmospheric search. As {\tt SYSREM} does not distinguish between the telluric and stellar lines and removes them both simultaneously, measurements of stellar RVs are therefore not feasible. With regard to the planetary atmospheric search, since HD\,332231\,b has a relatively small $K_{\textrm{p}}$, meaning that any potential planetary atmospheric absorption signal does not shift significantly from one frame to the next, a significant portion of such signal would most likely also be removed by {\tt SYSREM} along with the stationary stellar and telluric lines.

\section{Analysis}
\label{sec: Analysis}

\subsection{RV calculation}
\label{sec: RV calculation}
\begin{figure}
    \centering
    \includegraphics[width=\linewidth]{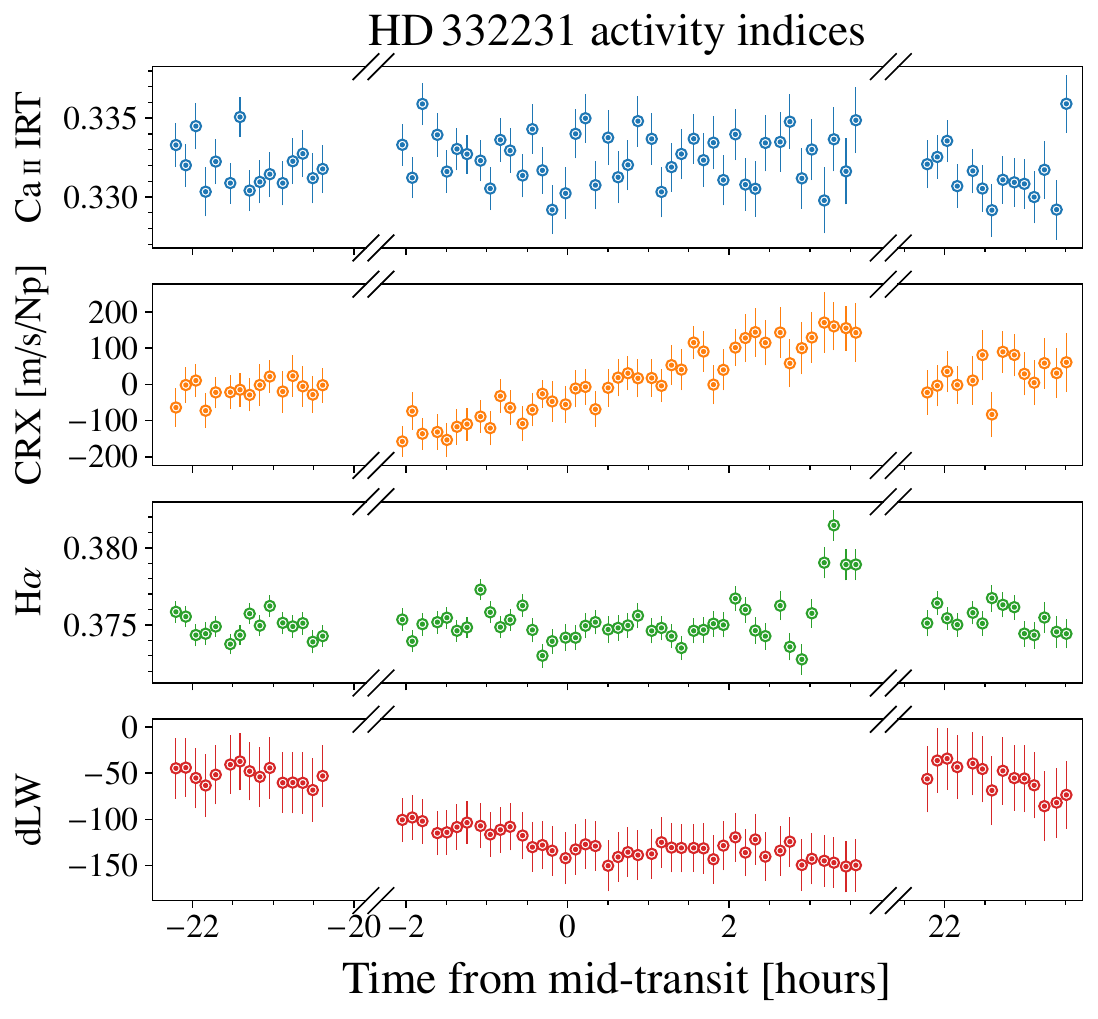}
    \caption{Stellar activity diagnostics calculated by {\tt serval}. From top to bottom: Ca\,{\scriptsize II} IRT index calculated from the middle line of the triplet at $\pm$15\,km/s; the chromatic RV index, which represents the dependence of RV on wavelength and is clearly affected by the transiting planet; H\textalpha~activity index calculated for bin a of $\pm$40\,km/s centred on the line core; and the differential Line Width again impacted by line deformation caused by the transiting planet. Overall the star presents no significant sign of variability due to activity.}
    \label{fig: Diagnostics}
\end{figure}

\begin{figure}
    \centering
    \includegraphics[width=\linewidth]{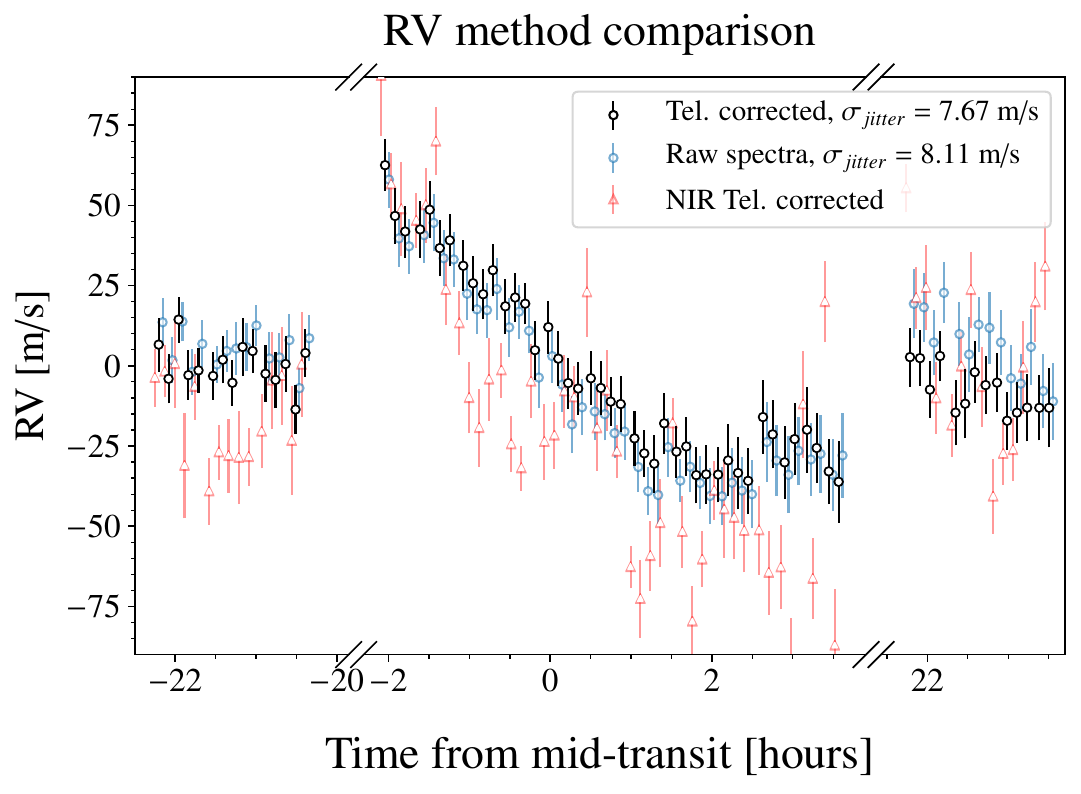}
    \caption{HD\,332231 RVs measured during the exoplanet transit, as well as pre- and post-transit epochs. The two VIS sets of measurements are made for the raw pipeline-corrected spectra (blue) and telluric-corrected spectra (black). The raw spectra have been shifted along the time axis for clarity. Additionally, the RVs measured from the NIR channel data have also been plotted in the background as red triangles, which have also been shifted along the time axis in the opposite direction.}
    \label{fig: RV comparison}
\end{figure}

We obtained the RV values of the star for the full set of observations using the {\tt serval} pipeline \citep{Zechmeister2018}, applying it to both sets of raw and telluric corrected spectra from the VIS and NIR channels. This pipeline calculates the RVs through a template matching approach \citep{Butler1996,Zechmeister2018}, which it constructs by coadding all available spectra for the target. It provides values which are corrected for instrumental drift, as well as secular acceleration. Additionally, the pipeline provides a whole host of spectral diagnostics, some of which are plotted in Fig. \ref{fig: Diagnostics}. Further to the pipeline calculated values, we also measured the RVs by fitting a Gaussian peak to the cross correlation function \citep[CCF;][]{Pepe2002} of the observed spectra with a {\tt phoenix} stellar model \citep{Husser2013} matching best the parameters of HD\,332231 and observed no statistically significant differences between the two sets of results (all pairs of values within 1\textsigma). Subsequently, throughout this analysis we present RV results calculated by the pipeline through the template-matching approach. These calculated RV values are presented in Fig. \ref{fig: RV comparison}, whereby a comparison between the values estimated from the raw and telluric corrected spectra is made. Considering only the out of transit spectra, the RV jitter for the target reduces by $\sim$\,0.44\,m/s; namely reduced from 8.11\,m/s for the RVs measured from the raw spectra to 7.67\,m/s for those estimated from the telluric corrected spectra. The improvements made to stellar RV measurements through telluric correction of high resolution spectra have previously been demonstrated in a number of studies \citep{Artigau2014,Cunha2014,Figueira2016,Kimeswenger2021}. The calculated RV values from both sets of spectra are given in Table \ref{tab: RVs}. Furthermore, we ignored the RV values determined from the NIR channel, which presented significantly larger uncertainties and jitter as compared to the VIS channel data (c.f. Fig. \ref{fig: RV comparison}). This is rather expected due to the lower S/N spectra in this channel owing to the spectral type of the star, as well as the presence of significant telluric absorption residuals and sky emission lines.

\subsection{Obliquity measurement}
\label{sec: Obliquity}
\begin{figure}
    \centering
    \includegraphics[width=\linewidth]{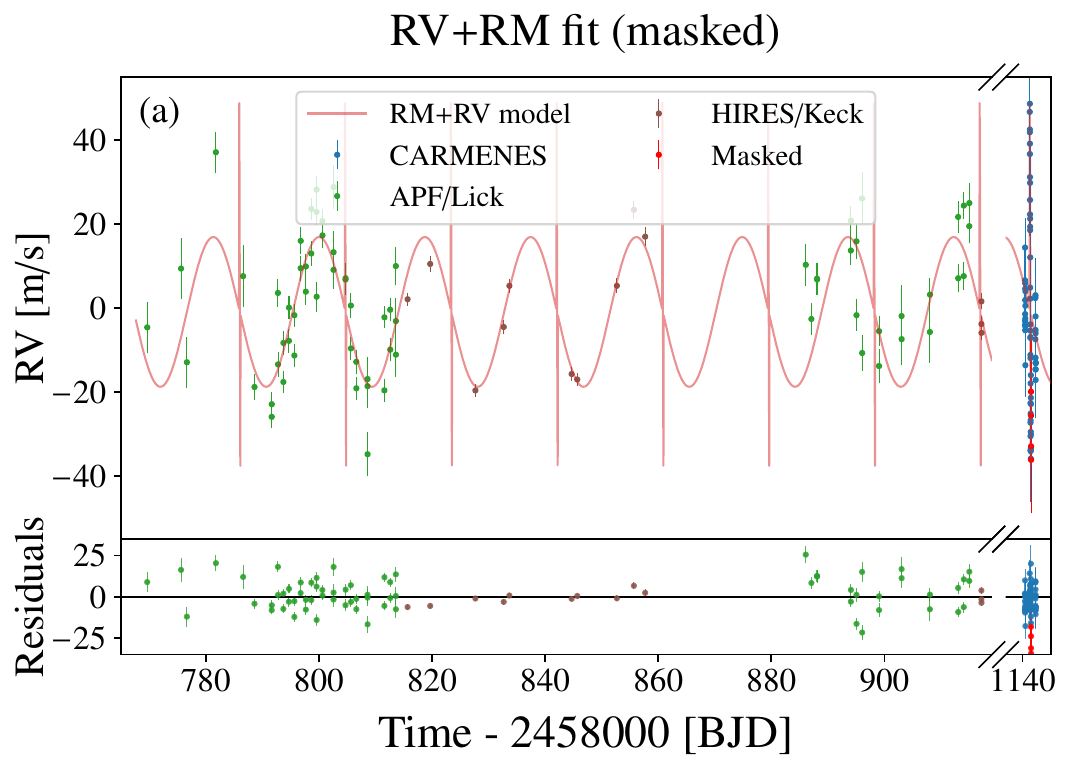}
    \includegraphics[width=\linewidth]{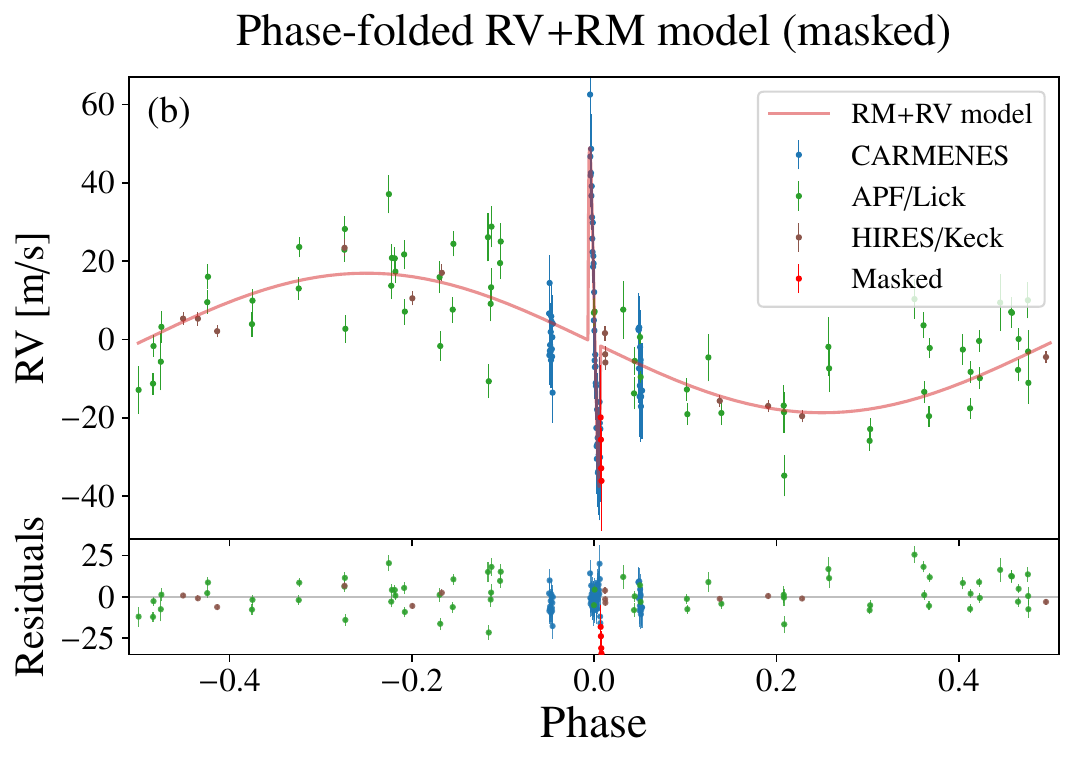}
    \includegraphics[width=\linewidth]{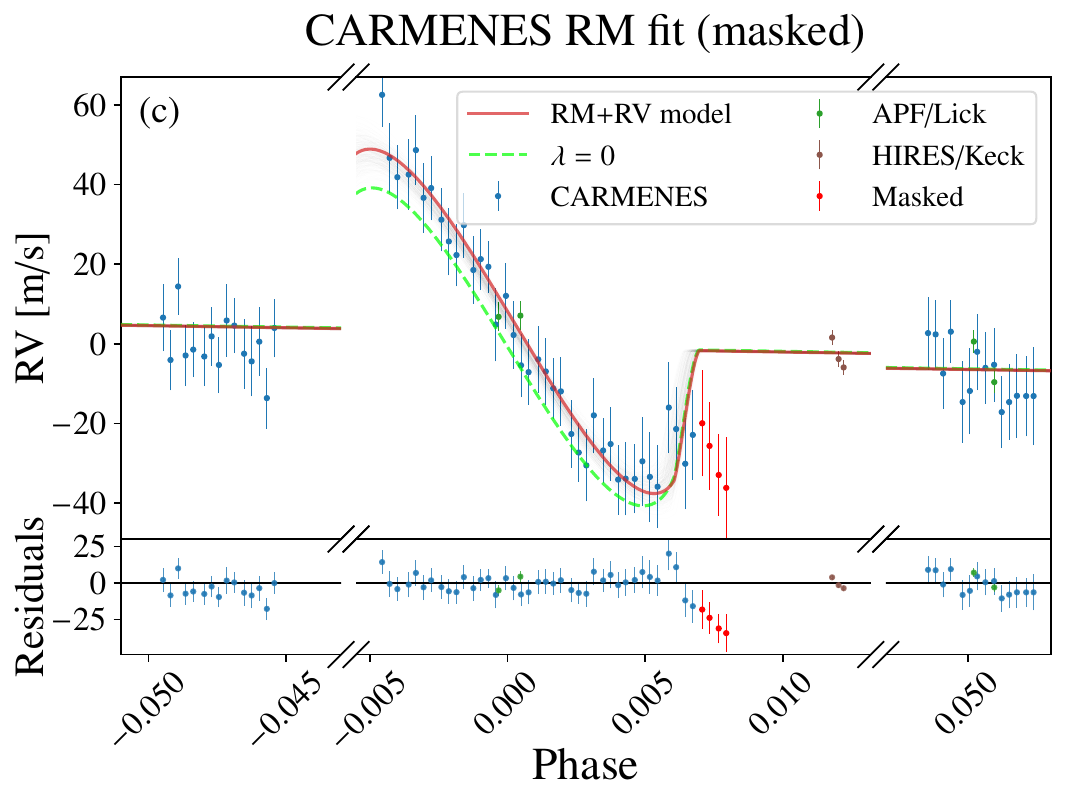}
    \caption{Joint RM+RV analysis of all available RV values for HD\,332231, excluding the last 4 out of transit points of CARMENES 18-10 observations shown as red points. (a) shows the joint analysis of previously obtained RV data from spectrographs at Keck (HIRES) and Lick (APF) observatories \citep{Dalba2020}, together with the data obtained with CARMENES. (b) represents the same data phase-folded for our derived system parameters from the fit above. (c) shows a zoom into the epoch of CARMENES observations (including the planetary transit) and the fitted RM function to the data. The thin grey lines are 300 random realisations drawn from 1\textsigma~regions of the posteriors. The lime dashed line represents the best fit model when \obli is fixed to zero, performed for the purpose of robustness estimation. The residuals of this model are omitted for clarity.}
    \label{fig: RV fit}
\end{figure}

\begin{figure*}
    \centering
    \includegraphics[width=0.7\linewidth]{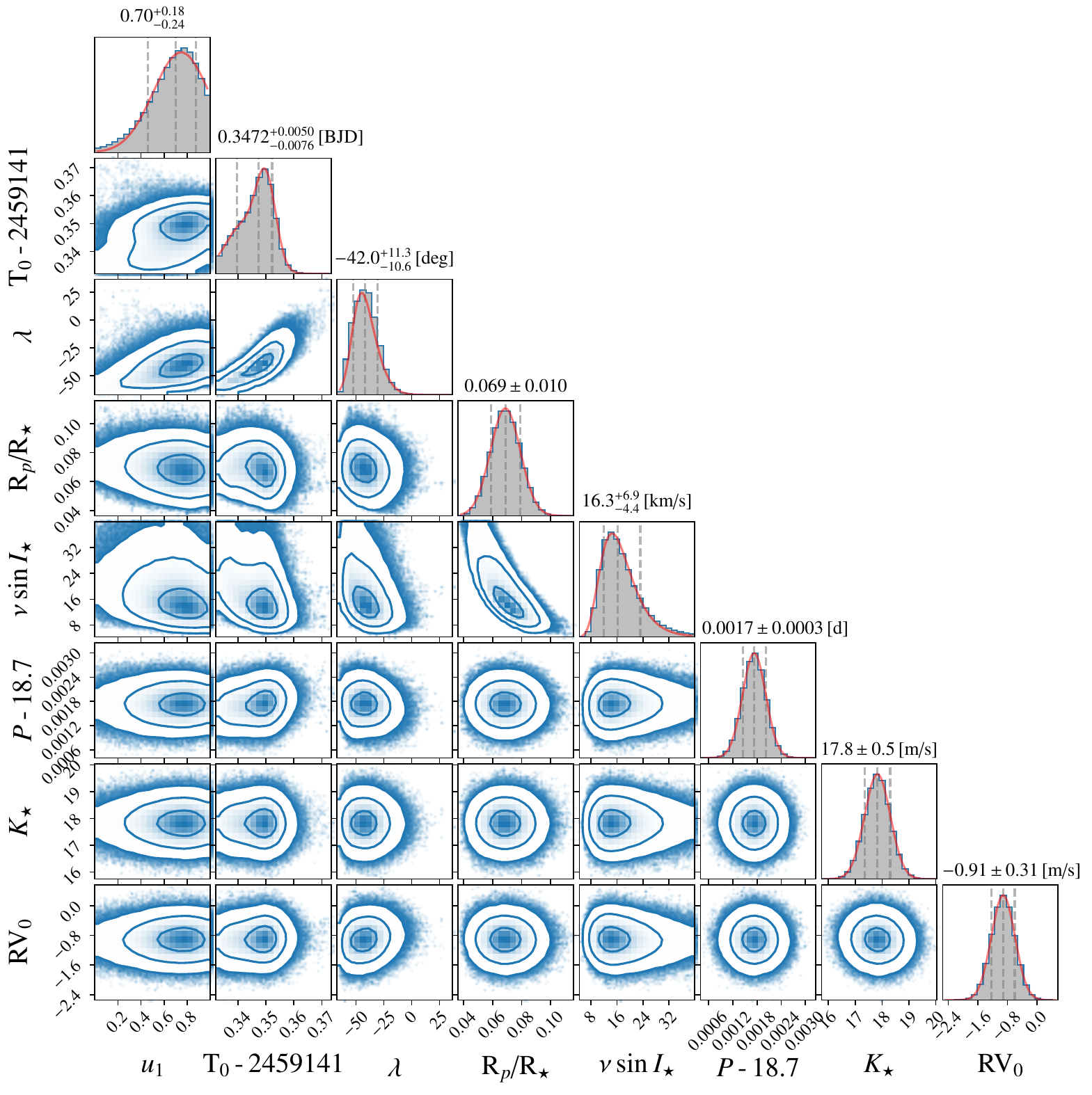}
    \caption{Posterior distributions for the fitted parameters of the RV+RM model. The results are for data where the last 4 data points of CARMENES 18-10 observations have been masked. The median values are quoted as solutions, with the 16th and 84th quartiles given as uncertainties at the top of each column, which are represented as vertical grey dashed lines. All posteriors are modelled well with a Gaussian function with the exception of the mid-transit time, where a double-degenerate solution is derived, as well as \obli and $\nu\sin I_\star$, where slightly skewed normal distributions are used. In the parameter-parameter plots, 1-, 2- and 3\textsigma~levels are presented as solid blue lines.}
    \label{fig: posteriors}
\end{figure*}

In order to possibly refine the orbital parameters obtained by \citet{Dalba2020}, as well as model the RM effect, we defined a composite Keplerian model for a circular orbit of a transiting planet, namely one that also includes the RM effect. We assumed a circular orbit as \citet{Dalba2020} obtained a value for eccentricity consistent with $\sim$\,0 ($0.032^{+0.030}_{-0.022}$). This was performed using the {\tt modelSuite} sub-routine from the {\tt PyAstronomy} python package \citep{Czesla2019}, which employs the formulation of \citet{Ohta2005} for the analytical description of the RM effect. In order to estimate best fitting values for the free parameters, we ran Markov Chain Monte Carlo (MCMC) simulations with 10$^6$ steps, burning in the first 10$^5$, through which we sampled from the posterior probability distributions. The best fit model, fitted to the three data sets used (HIRES/Keck, APF/Lick \citep{Dalba2020} and CARMENES), is shown in panel (a) of Fig. \ref{fig: RV fit}, with the phase-folded RV curve shown in panel (b) and a zoom into the in-transit region, showing clearly the RM effect and fitted model, in panel (c) of the same figure. The best fit parameters and their uncertainties, the prior assumptions, as well as a comparison to previously published values are given in Table \ref{tab: RV+RM fit results}. We note that the last 4 observations performed during the night of 18-10, which are out of transit, present large and significant residuals for the fitted RM model. The nature of this offset is not entirely clear and it is most likely due to the fact that they were taken for the star at very high airmass ($1.75 \rightarrow 1.98$). In order to investigate whether these data points bias any of the results for the fitted parameters, we performed the analysis twice; once with all the data included and a second time with these data points masked. The results from fitting data with masked out of transit points are given in Figs. \ref{fig: RV fit} and \ref{fig: posteriors} and their equivalents with all the data points included in Figs. \ref{fig: RV fit appx} and \ref{fig: posteriors appx}. A comparison between the posterior distributions shows that the inclusion of these 4 data points only marginally impacts the final outcome and the results are not sensitive to them in any statistically significant way.

For the fitting of the model, we fixed the scaled semi-major axis ($a/R_\star$) and the inclination of the planetary orbit ($i$) to those values found by \citet{Dalba2020} from fitting the TESS transit light curves with an analytical model. This method provides much more stringent solutions as compared to what could be obtained from fitting the RM effect. Furthermore, the inclination of the stellar rotation axis ($I_\star$) is also fixed to 90\,$\deg$, since leaving it as a free parameter did not result in convergence of the posterior. This fact, of course, means that our fitted value for the projected rotational velocity of the star ($\nu\sin I_\star$) is only an upper limit. All other parameters were taken as free, with the assumed prior distributions, as well as the best fit values derived from the analysis of posterior probabilities, given in Table \ref{tab: RV+RM fit results}. These posterior probability distributions are plotted in Fig. \ref{fig: posteriors}.

\renewcommand{\arraystretch}{1.5}
\begin{table*}[]
    \centering
    \caption{Orbital and physical parameters of HD\,332231\,b system from simultaneous RV and RM modelling of all data available, excluding those masked.}
    \begin{tabular}{l c c c}
    \hline \hline
    Parameter & Prior & This study & \citet{Dalba2020} \\
    \hline
    Period [d]      & $\mathcal{U}$(P$_{\mathrm{D}}-\frac{1}{24}$, P$_{\mathrm{D}}+\frac{1}{24}$)     & $18.70173 \pm 0.00028$ & $18.71204 \pm 0.00043$ (P$_{\mathrm{D}}$) \\
    T$_0 \mathbf{- 2459141}$ [BJD]     & $\mathcal{U}$(T$_{\mathrm{D}}-\frac{0.5}{24}$, T$_{\mathrm{D}}+\frac{0.5}{24}$) & $0.34721^{+0.00497}_{-0.00764}$ & 0.35287 (T$_{\mathrm{D}}$) \\
    $R_{\textrm{p}}$/$R_\star$ & $\mathcal{N}$(0.06976, 0.01$^2$)& $0.06893^{+0.01027}_{-0.00998}$ & $0.06976^{+0.00041}_{-0.00039}$\\
    \obli [$\deg$]           & $\mathcal{U}$($-90$, 90) & $-42.0^{+11.3}_{-10.6}$ & ... \\
    $\nu \sin I_\star$ [km/s] & $\mathcal{U}$(0, 40)     & $16.3^{+6.9}_{-4.4}$\tablefootmark{a} & ($5.3 - 7.0$) $\pm 1.0$ \\
    $K_\star$ [m/s]            & $\mathcal{U}$(10, 30)    & $17.8 \pm 0.5$ & $17.3 \pm 1.2$\\
    RV$_0$ [m/s]               & $\mathcal{U}$($-10$, 10)   &$-0.91 \pm 0.31$         & ... \\ 
    $u_1$                      & $\mathcal{U}$(0, 1)      & $0.70^{+0.18}_{-0.24}$ & ...\tablefootmark{b} \\
    $a/R_\star$                & ... & $<24.21>$ & $24.21^{+0.62}_{-0.78}$\\
    $i$ [$\deg$]             & ... & $<89.68>$ & $89.68^{+0.22}_{-0.28}$\\
    $I_\star$ [$\deg$]       & ... & $<90>$    & ... \\
    $M_p$ [$M_{\mathrm{Jup}}$]              & ... & $0.251 \pm 0.017$   & $0.244 \pm 0.021$  \\
    $\rho_p$ [g\,cm$^{-3}$]    & ... & $0.478^{+0.047}_{-0.044}$& $0.464^{+0.054}_{-0.052}$\\
    \hline
    \end{tabular}
    \tablefoot{For the period and mid-transit time, values from \citet{Dalba2020} are used to construct prior distributions. $<>$ symbols indicate fixed parameters.
    \tablefoottext{a}{This is an upper limit for the sky-projected stellar rotation velocity and cannot directly be compared to the values from \citet{Dalba2020}, as they are derived from spectral analysis.}
    \tablefoottext{b}{In fitting the transit light curve, \citet{Dalba2020} implement the quadratic limb darkening law, whereas the analytical RM model of \citet{Ohta2005} employs the linear formulation.}}
    \label{tab: RV+RM fit results}
\end{table*}

From this analysis we obtained a value of $-42.0^{+11.3}_{-10.6}$\,$\deg$ for the orbital obliquity angle (\textlambda) of HD\,332231\,b, with $\nu \sin I_\star = 16.3^{+6.9}_{-4.4}$\,km/s. The value of \obli has been plotted in the four parameter spaces presented earlier in Fig. \ref{fig:Obliquity_dist}. The best fit RM model to the in-transit data has been highlighted in panel (c) of Fig. \ref{fig: RV fit}, where 300 random realisations from the inner 1-\textsigma~posterior distributions have also been drawn, representing the uncertainty in the final fitted model.

We note that there exists a degeneracy between the mid-transit time (T$_0$) and the obliquity angle (\textlambda), as is evident from their mutual posterior plot in Fig. \ref{fig: posteriors}. Ideally one would fix the mid-transit time to the value derived from the ephemeris obtained from photometry as that method measures this value with a much higher precision than the RM analysis. However, since there are 24 orbits of the planet around its host star between the epoch given by \citet{Dalba2020} and the one observed for this study, the uncertainty in the predicted mid-transit time in our CARMENES observations are rather large. This uncertainty is approximately taken as the extremes of the uniform prior distribution assumed in our fitting analysis given in Table \ref{tab: RV+RM fit results}. To assess the impact that this apparent correlation has on the determined obliquity value, we also ran the analysis with the mid-transit time fixed to the one predicted from TESS ephemeris. This approach led to an obliquity value of $-44.8^{+5.3}_{-4.6}$\,deg, indicating that the present degeneracy does not significantly alter the conclusion of moderate misalignment.

\subsection{Transmission spectroscopy}
\label{sec: transmission spectroscopy}
We searched for possible atmospheric signatures from individually resolved lines emanating from the exoplanet. This is performed through two distinct, albeit fundamentally equivalent, approaches, namely; 1) searching for absorption from strong singular transition lines from species such as Na, K, H or He, or 2) systematically adding absorption signals from a multitude of weaker lines through the cross correlation technique.

\begin{figure*}
    \centering
    \includegraphics[width=\linewidth]{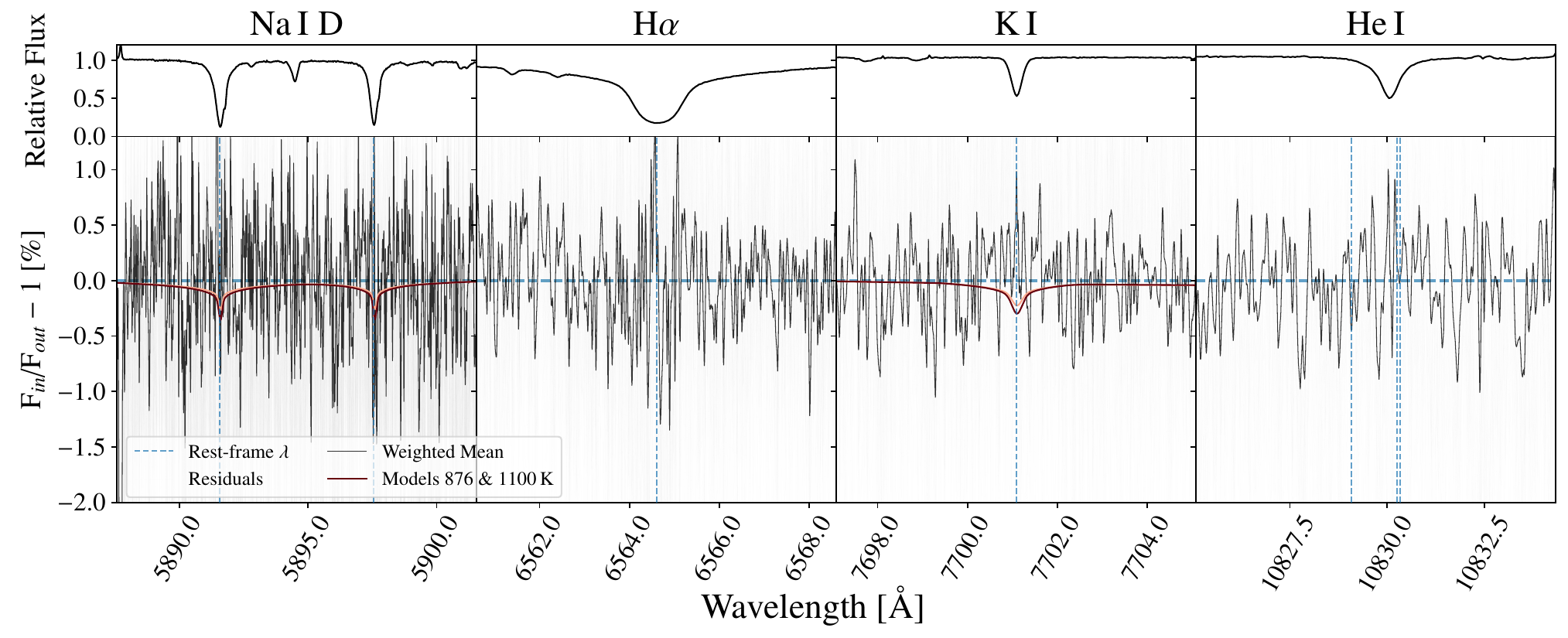}
    \caption{Narrowband transmission spectroscopy of strong transition lines. The upper panels present the composite out of transit stellar spectrum. We note that, with the exception of He\,{\scriptsize I}, all the absorption lines present in the stellar spectrum correspond to the species being probed, labelled above each column. The bottom panels present the transmission spectra. The thin grey lines are the individual in-transit residuals shifted to the planetary rest frame. The black lines are the weighted mean of all those shifted residuals. The rest-frame wavelength of each individual transition probed are indicated with vertical blue dashed lines. For Na and K, two atmospheric models of solar metallicity and C/O, with equilibrium temperatures of 876 and 1100\,K have been over plotted.}
    \label{fig: TS narrowband}
\end{figure*}

The first of the two aforementioned methods has previously been employed in a number of studies to detect a host of atomic species in the upper atmospheres in a multitude of transiting exoplanets. An up to date list of these detected species through narrowband transmission spectroscopy is given in the Appendix Table \ref{tab: Narrowband detections}. We performed this analysis of HD\,332231b for all four of these neutral species, the results for which are shown in Fig. \ref{fig: TS narrowband}. It is quite evident that the precision at which the data probes the atmosphere is far below what is needed for any possible detection in this atmosphere due to its relatively low equilibrium temperature. Furthermore, a comparison to atmospheric models suggests that placing any upper limits on the temperature or abundances from these data is not particularly informative. These models are calculated with the {\tt petitRADTRANS} package \citep{Molliere2019}, with equilibrium chemistry considered through {\tt FastChem} \citep{Stock2018}, examples of which are shown in two panels of Fig. \ref{fig: TS narrowband}.

\begin{figure}
    \centering
    \includegraphics[width=\linewidth]{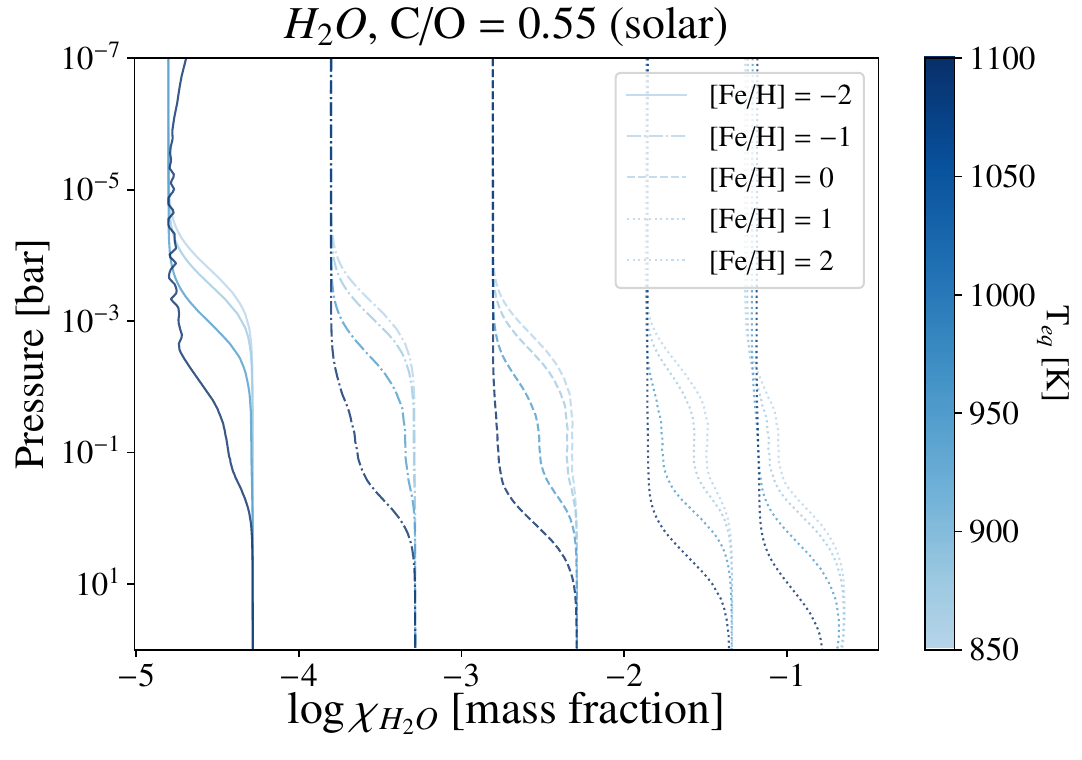}
    \caption{H$_2$O abundance profiles in mass fraction, \textchi. These are calculated for HD\,332231\,b, for a range of atmospheric equilibrium temperatures, $825 \leq T_{eq} \leq 1100$\,K and metallicities, $-2 \leq \textrm{[Fe/H]} \leq 2$, with C/O fixed to the solar value of 0.55. These profiles, determined with {\tt FastChem}, were subsequently used to calculate the models injected into the data, retrieval of which yields the detection space presented in Fig. \ref{fig: CCF H2O}. }
    \label{fig: H2O abundances}
\end{figure}

In addition to this narrow-band approach, we searched for the possible presence of H$_2$O in the atmosphere through the addition of absorption signals from a multitude of individually resolved lines in various strong bands of water in the visible and the near-IR wavelengths. This was performed through the calculation of the CCF of the telluric-corrected spectra in the stellar rest frame with atmospheric templates. The weighted CCF is defined as:
\begin{equation}
       C(v,t) = \frac{\sum_i^N x_i(t) ~ T_i(v)}{\sum_i^N T_i(v)}
\end{equation}
\noindent where $T(\nu)$ is the atmospheric template Doppler shifted to velocity $\nu$, and $x(t)$ is the residual planet spectrum observed at time $t$. We note that in order to perform the operation in the numerator, both $T$ and $x$ must necessarily be sampled along the same wavelength grid, which is typically achieved by calculating the template at very high resolution and re-sampling it onto the observational wavelength grid. The cross correlation $C$ is subsequently a two-dimensional matrix with rows representing epochs of observations and the columns spanning the velocity space being probed. During this analysis, for a slightly more comprehensive assessment of data quality and detection limits, we calculate a grid of models varying in equilibrium temperature and metallicity and inject them into the residual planet spectra through multiplication. The abundance profiles for this range of models calculated with the equilibrium chemistry code {\tt FastChem} are presented in Fig. \ref{fig: H2O abundances}.

\begin{figure}
    \centering
    \includegraphics[width=\linewidth]{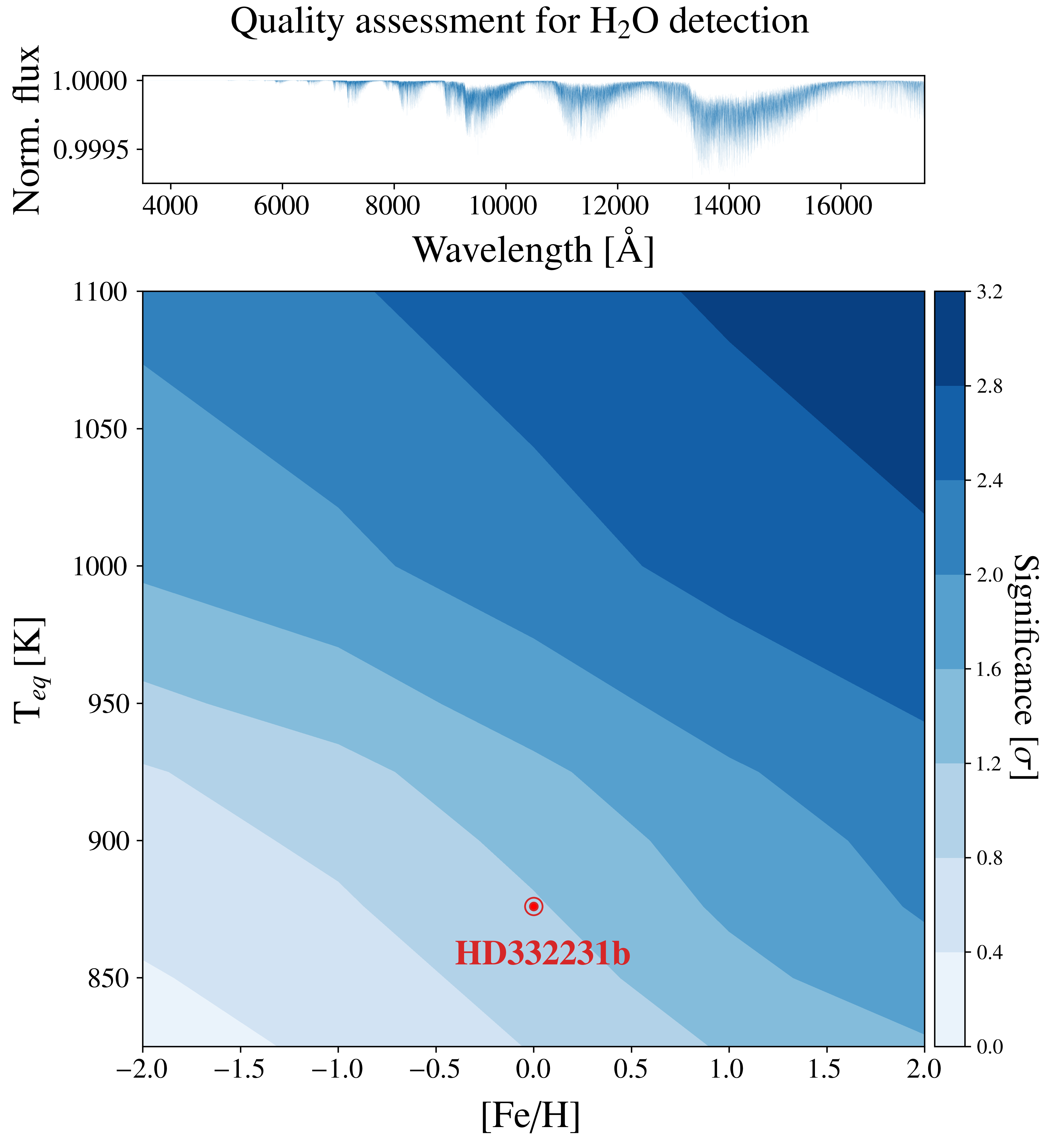}
    \caption{Assessing the quality of CARMENES observations for the potential detection of H$_2$O in the atmosphere of HD\,332231\,b. \textit{Top} panel shows an example of a transmission model calculated with {\tt petitRADTRANS} (T$_{eq}$\,=\,876\,K and [Fe/H]\,=\,0), normalised to a baseline estimated from the continuum species only and convolved with the instrumental profile. \textit{Bottom} panel shows the detection significances for the injected models, smoothed with a Gaussian filter of \textsigma\,=\,1. The expected location of HD\,332231\,b's atmosphere for its estimated equilibrium temperature and assumed solar metallicity has been indicated as the red data point.}
    \label{fig: CCF H2O}
\end{figure}

The cross correlation maps including the injected signals were then systematically summed for a range of planetary radial velocities to obtain the so-called velocity-velocity maps from which detection significances are determined. We performed this analysis for H$_2$O only, taking advantage of the deep absorption bands in the near-IR wavelengths, as shown in the top panel of Fig. \ref{fig: CCF H2O}. The {\tt FastChem} equilibrium chemistry model for the calculated planetary $T_{eq}$, solar C/O and metallicity, for a wide range of species is given in Fig. \ref{fig: FastChem abundances}. 

As mentioned above, atmospheric models, calculated using {\tt petitRADTRANS} and {\tt FastChem}, spanning in temperature and metallicity were injected into the data and their corresponding peaks in the velocity maps were compared to the variance of the entire map to determine the significance at which each point in the parameter space was detected. A smoothed map of this detection space is shown in the main panel of Fig. \ref{fig: CCF H2O}, where it is evident, given the quality of the data, that for the expected equilibrium temperature of the planet any presence of H$_2$O in the atmosphere would not be detectable. In fact much higher temperatures and abundances are necessary for any potential signal to be accessible. One important caveat that we note with this analysis is that we have assumed a C/O ratio equal to that of the solar value, as well as a cloudless and symmetric atmosphere in hydrostatic and chemical equilibrium. All such assumptions, of course, have impacts upon the detectability space to varying degrees.

\section{Discussion}
\label{sec: Discussion}
\subsection{Robustness of \obli measurement}
From the analysis of the RV values measured for HD\,332231, we presented a moderately misaligned orbit for the warm sub-Saturn planetary companion of this star. This obliquity angle was measured through the fitting of an RV+RM model to all the available data points, whereby the in-transit data were almost exclusively obtained through our observations with CARMENES. This angle was measured at \obli$=-42.0^{+11.3}_{-10.6}$\,$\deg$, which in the context of the relatively small, warm planet sample (presented in panel (b) of Fig. \ref{fig:Obliquity_dist}), presents a rather rare moderately misaligned orbit.

In order to assess the significance of this measurement, we performed a statistical comparison between the currently fitted model and one in which \obli is fixed to 0\,$\deg$. We fitted all the available data again with an identical procedure to the one described in section \ref{sec: Obliquity}, whereby the only difference was that the planetary orbit is assumed to be completely aligned with the stellar rotation axis (\obli = 0). The result of this analysis is shown in panel (c) of Fig. \ref{fig: RV fit}, as the lime dashed line. We then compare these two models to the data obtained with CARMENES only (excluding the masked points) to determine the significance of fit improvement by keeping \obli free. Making the approximation that the residuals of both models are normally distributed, we calculate the likelihood ratio, \textit{LR}, as:
\begin{equation*}
    LR = 2 \ln{\left( \mathcal{L}(\theta_{\lambda=0}) - \mathcal{L}(\theta_{\lambda=free}) \right)} = 14.38
\end{equation*}
\noindent where $\mathcal{L}(\theta)$ are the log likelihood values of the models with \obli fixed and taken as a free parameter. This ratio corresponds to $p=0.00015$, with the critical value at $P=0.005$ for the degree of freedom 1 being 7.879. Therefore, we reject the null hypothesis (H$_0$: no statistically significant difference between the variances of the residuals) at high significance and conclude that the model with free \obli describes best the data. The F-ratio, which in this case is a monotone transformation of the likelihood ratio, also results in an identical conclusion.

\subsection{Tidal effects}
\citet{Albrecht2012} estimate the realignment time scale for planets with host stars with convective envelopes ($\tau_{CE}$) as:
\begin{equation}
    \frac{1}{\tau_{CE}} = \frac{1}{10^{10} \textrm{yr}} \left( \frac{M_p}{M_\star} \right)^2 \left( \frac{40}{a/R_\star} \right)^6
\end{equation}
\noindent which is estimated through a calibration of relations derived by \citet{Zahn1977}, with binary star observations. We chose to use this relation, as opposed to its equivalent for those stars with radiative envelopes, because HD\,332231 is a main sequence star with a measured effective temperature of $6089^{+97}_{-96}$\,K, in other words spectral type F8 with a radiative core and a convective envelope. This time scale for HD\,332231\,b is therefore estimated as $\sim$\,$10^{16}$\,yr, far exceeding the age of the universe. Subsequently, assuming that the host star has mainly a convective envelope, tidal interactions with the host likely have played a minimal role in altering the obliquity of the planetary orbit.

However, given that the stellar temperature is relatively close to the Kraft break limit \citep[$\sim$\,6200\,K;][]{Kraft1967}, where the transition from convective to radiative envelopes is expected, we also considered the realignment time scale for the radiative regime. \citet{Zahn1977} estimate this as:
\begin{equation}
    \frac{1}{\tau_{RE}} = \frac{4}{5 \times 10^9 \textrm{yr}} \left( \frac{M_p}{M_\star} \right) ^2 \left( 1+ \frac{M_p}{M_\star} \right)^{5/6} \left( \frac{a/R_\star}{6} \right) ^{-17/2}
\end{equation}
\noindent which for the HD\,332231 system is estimated as $\sim$\,$10^{21}$\,yr. Therefore, if the star in fact does have a significant radiative envelope, the tidal effects are expected to be even less significant.

\subsection{Misalignment due to a hypothetical external companion}
\citet{Rice2021} recently suggested that giant planets around cooler stars, those with convective envelopes and therefore below the Kraft limit, tend to be on more misaligned orbits as compared to those orbiting hotter stars with radiative envelopes. Such trend, if indeed observed for a large sample, could be indicative of mechanisms such as those proposed by \cite{Anderson2018}. They suggest that the obliquity of a giant planet could be excited by an external, modestly inclined companion, due to a secular resonance that occurs when the precession rate of the stellar spin axis is similar to the nodal precession rate of the inner planet.

\begin{figure}
    \centering
    \includegraphics[width=\linewidth]{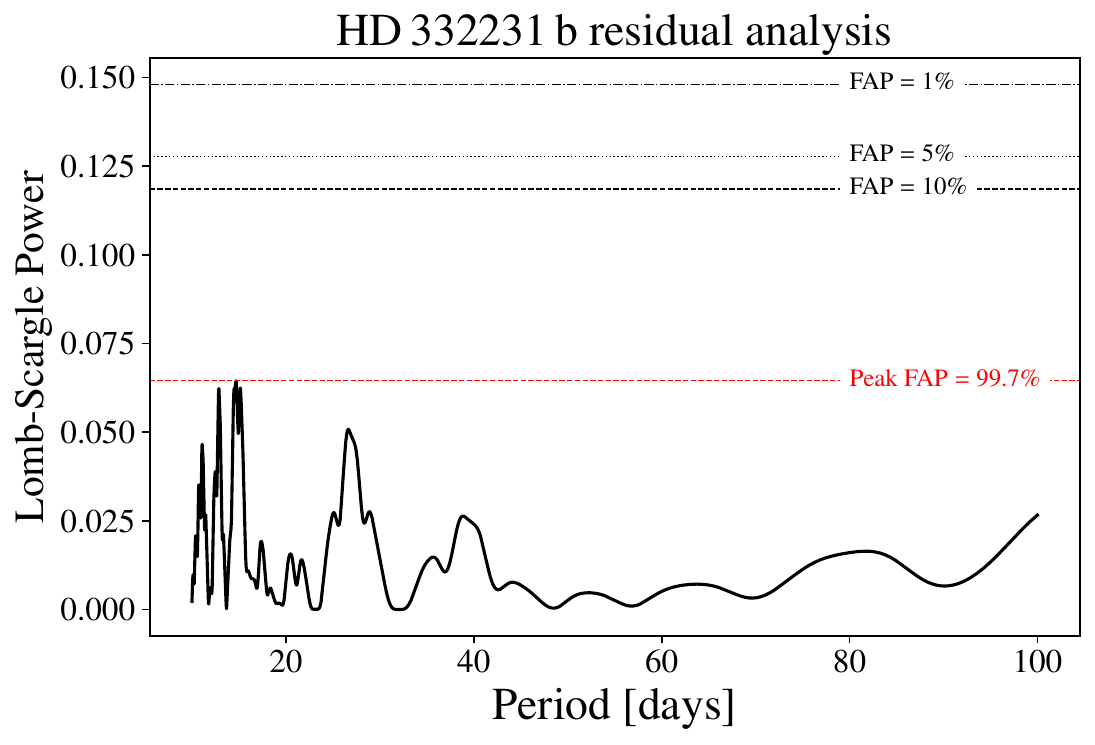}
    \caption{Lomb-Scargle periodogram of the HD\,332231 RV residuals for the RV+RM model of the warm-Saturn planet detected from TESS photometry. 1, 5 and 10\% False Alarm Probability (FAP) thresholds have been indicated. We do not detect any statistically significant periodic signal in the residuals. The red dashed line shows the FAP for the maximum peak identified in the periodogram.}
    \label{fig: residual LS}
\end{figure}

To investigate such a possibility, we searched for the RV signal of a hypothetical outer companion in the residuals of our model shown in panel (a) of Fig. \ref{fig: RV fit}. We did not however detect any statistically significant signal in the current data, the periodogram for which is shown in Fig. \ref{fig: residual LS}. Higher precision RV measurements, with a longer baseline are required to definitively confirm or reject the presence of an outer companion in this system, thereby testing the aforementioned hypothesis.

\section{Conclusions}
\label{sec: Conclusions}
In this study we presented high resolution spectroscopic transit observations of the warm sub-Saturn HD\,332231\,b, using the CARMENES spectrograph. The correction of the telluric absorption features using {\tt molecfit} improved the precision of the RV values obtained from those recorded spectra. Measurements of various activity diagnostics indicated that the observations are not likely to be affected by intrinsic variability of the star, which was further corroborated by the stability of the long term photometry of the star obtained by TESS \citep{Dalba2020}.

We measured a projected spin-orbit obliquity angle (\textlambda) of $-42.0^{+11.3}_{-10.6}$\,$\deg$ for HD\,332231\,b, which presents a moderately misaligned planet, with a relatively large orbital period. In the context of planet formation theory presented in section \ref{sec: Introduction}, such finding would suggest that those formation theories not invoking planetary migration are favoured. However, for such conclusions to be made concrete, certainly a much larger sample of obliquity measurements for this class of warm transiting planets is required. This is, of course, made rather difficult by the rarity and duration of the transits of those planets. We also discussed the robustness of the measured obliquity value through a statistical model comparison and showed that the inferred measurement significantly improves the residuals of the fitted model, as compared to an aligned orbit. Trivially, it was shown that tidal effects of the host star have not played a role in altering the alignment of the planetary orbital orientation. The residuals of the model were searched for the presence of an outer companion that could possibly excite the obliquity of the inner planet, which resulted in a non-detection.

As an aside, we also analysed the spectra for atmospheric signatures through a couple of distinct methods; namely narrowband transmission spectroscopy of strong, singular transition lines and cross correlation analysis with models containing a myriad of atmospheric lines of H$_2$O. We showed that given the precision of the obtained spectra, it is not feasible to detect any atmospheric characteristics of this planet, especially considering its relatively low estimated equilibrium temperature of 876K. Through an injection and retrieval routine, we presented a slice through the detection parameter space of the atmosphere for a range of metallicities and equilibrium temperatures. In order for such data to be amenable to atmospheric studies, higher precision instrumentation coupled to much larger aperture telescopes are required, a problem which spectrographs such as HIRES at the ELT \citep{Marconi2021} will address.

\begin{acknowledgements}
      CARMENES is an instrument at the Centro Astron\'omico Hispano-Alem\'an (CAHA) at Calar Alto (Almer\'{\i}a, Spain), operated jointly by the Junta de Andaluc\'ia and the Instituto de Astrof\'isica de Andaluc\'ia (CSIC). The authors wish to express their sincere thanks to all members of the Calar Alto staff for their expert support of the instrument and telescope operation.
      We acknowledge financial support from the Agencia Estatal de Investigaci\'on of the Ministerio de Ciencia, Innovaci\'on y Universidades through projects Ref. PID2019-110689RB-I00/AEI/10.13039/501100011033 and the Centre of Excellence ``Severo Ochoa'' award to the Instituto de Astrof\'isica de Andaluc\'ia (SEV-2017-0709). 
      We also acknowledge the use of the ExoAtmospheres database during the preparation of this work and thank the team at the IAC for their excellent work preparing such a useful database. ES acknowledges support from ANID - Millennium Science Initiative - ICN12\_009. KM acknowledges support from the Excellence Cluster ORIGINS, which is funded by the Deutsche Forschungsgemeinschaft (DFG, German Research Foundation) under Germany's Excellence Strategy - EXC-2094 - 390783311. A.S.L. acknowledges funding from the European Research Council under the European Union's Horizon 2020 research and innovation program under grant agreement No 694513.
      
      Beside those explicitly mentioned in the manuscript, throughout this work the following {\tt python} packages have been used: {\tt astropy} \citep{astropy}, {\tt corner} \citep{corner}, {\tt matplotlib} \citep{matplotlib}, {\tt NumPy} \citep{numpy}, {\tt pandas} \citep{pandas}, {\tt PyMC} \citep{pymc} and {\tt SciPy} \citep{scipy}. 
\end{acknowledgements}

%
%

\bibliographystyle{aa} 
\bibliography{HD332231.bib} 

\begin{appendix} 
\section{Calculated RVs}
RV values calculated by the {\tt serval} pipeline from both raw and telluric corrected spectra.

\renewcommand{\arraystretch}{1.2}
\begin{table}[h!]
\caption{{\tt Caracal} calculated RVs.}             
\label{tab: RVs}      
\centering              
  \begin{tabular}{l c c}
    \hline \hline
      Time [BJD]  & \multicolumn{2}{c}{RV} \\ \cline{2-3}
       & Raw spectra & Telluric-corrected \\
            & [m/s]       & [m/s] \\
    \hline
2459140.42205    &       ~13.57 $\pm$ 7.51        &       ~~6.61 $\pm$ 8.40   \\
2459140.42713    &       ~~1.74 $\pm$ 7.17        &       ~-4.04 $\pm$ 7.63   \\
2459140.43232    &       ~13.80 $\pm$ 6.18        &       ~14.42 $\pm$ 7.15   \\
2459140.43736    &       ~-1.89 $\pm$ 7.19        &       ~-2.90 $\pm$ 7.71   \\
2459140.44257    &       ~~6.84 $\pm$ 7.36        &       ~-1.43 $\pm$ 6.93   \\
2459140.45008    &       ~~0.41 $\pm$ 5.76        &       ~-3.16 $\pm$ 7.45   \\
2459140.45508    &       ~~4.62 $\pm$ 6.62        &       ~~1.90 $\pm$ 7.40   \\
2459140.46005    &       ~~5.41 $\pm$ 8.15        &       ~-5.29 $\pm$ 7.08   \\
2459140.46531    &       ~~5.83 $\pm$ 7.48        &       ~~5.88 $\pm$ 9.07   \\
2459140.47045    &       ~12.57 $\pm$ 6.42        &       ~~4.61 $\pm$ 6.91   \\
2459140.47712    &       ~~2.29 $\pm$ 8.34        &       ~-2.46 $\pm$ 8.81   \\
2459140.48229    &       ~~2.61 $\pm$ 7.70        &       ~-4.38 $\pm$ 8.78   \\
2459140.48746    &       ~~8.01 $\pm$ 8.23        &       ~~0.56 $\pm$ 8.60   \\
2459140.49260    &       ~-6.95 $\pm$ 6.87        &       -13.61 $\pm$ 7.66   \\
2459140.49770    &       ~~8.59 $\pm$ 7.15        &       ~~4.00 $\pm$ 7.36   \\
2459141.26203    &       ~58.04 $\pm$ 8.72        &       ~62.54 $\pm$ 8.00   \\
2459141.26715    &       ~39.72 $\pm$ 8.88        &       ~46.71 $\pm$ 8.71   \\
2459141.27239    &       ~37.27 $\pm$ 8.62        &       ~41.86 $\pm$ 8.10   \\
2459141.28010    &       ~40.74 $\pm$ 8.69        &       ~42.53 $\pm$ 8.83   \\
2459141.28502    &       ~44.57 $\pm$ 8.92        &       ~48.66 $\pm$ 8.88   \\
2459141.29021    &       ~33.54 $\pm$ 8.74        &       ~36.68 $\pm$ 8.72   \\
2459141.29538    &       ~33.16 $\pm$ 8.62        &       ~39.14 $\pm$ 8.12   \\
2459141.30233    &       ~22.48 $\pm$ 8.22        &       ~31.19 $\pm$ 7.93   \\
2459141.30736    &       ~17.60 $\pm$ 7.72        &       ~25.73 $\pm$ 8.50   \\
2459141.31257    &       ~17.35 $\pm$ 8.54        &       ~22.31 $\pm$ 7.87   \\
2459141.31760    &       ~23.95 $\pm$ 9.73        &       ~29.82 $\pm$ 8.17   \\
2459141.32402    &       ~12.00 $\pm$ 9.16        &       ~18.55 $\pm$ 8.53   \\
2459141.32908    &       ~16.88 $\pm$ 8.37        &       ~21.29 $\pm$ 7.66   \\
2459141.33426    &       ~10.93 $\pm$ 7.09        &       ~19.36 $\pm$ 6.48   \\
2459141.33930    &       ~-3.67 $\pm$ 9.50        &       ~~4.87 $\pm$ 9.23   \\
2459141.34620    &       ~~3.01 $\pm$ 8.30        &       ~12.06 $\pm$ 8.25   \\
2459141.35135    &       ~-5.85 $\pm$ 8.45        &       ~~2.23 $\pm$ 8.47   \\
2459141.35644    &       -18.26 $\pm$ 8.94        &       ~-5.37 $\pm$ 7.96   \\
2459141.36158    &       -12.92 $\pm$ 8.68        &       ~-7.08 $\pm$ 8.23   \\
2459141.36812    &       -14.23 $\pm$ 8.86        &       ~-3.88 $\pm$ 8.40   \\
2459141.37326    &       -15.08 $\pm$ 8.14        &       ~-6.91 $\pm$ 8.37   \\
2459141.37834    &       -21.00 $\pm$ 7.86        &       -11.15 $\pm$ 7.60   \\
2459141.38351    &       -20.50 $\pm$ 8.74        &       -11.91 $\pm$ 8.63   \\
2459141.39060    &       -31.54 $\pm$ 7.81        &       -22.62 $\pm$ 8.45   \\
2459141.39564    &       -39.09 $\pm$ 7.41        &       -27.28 $\pm$ 7.36   \\
2459141.40081    &       -40.27 $\pm$ 8.92        &       -30.50 $\pm$ 9.05   \\
\end{tabular}
\end{table}

\setcounter{table}{0}
\renewcommand{\thetable}{A.\arabic{table}}
\renewcommand{\arraystretch}{1.2}
\begin{table}[h!]
\caption{\textit{continued.}}             
\centering              
  \begin{tabular}{l c c}
    \hline \hline
      Time [BJD] & \multicolumn{2}{c}{RV} \\ \cline{2-3}
       & Raw spectra & Telluric-corrected \\
            & [m/s]       & [m/s] \\
    \hline
2459141.40591    &       -25.35 $\pm$ 9.21        &       -17.93 $\pm$ 9.42   \\
2459141.41225    &       -35.80 $\pm$ 6.61        &       -26.76 $\pm$ 8.08   \\
2459141.41732    &       -31.45 $\pm$ 7.90        &       -25.13 $\pm$ 9.21   \\
2459141.42243    &       -36.51 $\pm$ 8.27        &       -34.07 $\pm$ 8.79   \\
2459141.42755    &       -40.56 $\pm$ 8.60        &       -33.83 $\pm$ 9.21   \\
2459141.43373    &       -40.67 $\pm$ 9.00        &       -33.90 $\pm$ 8.68   \\
2459141.43889    &       -36.43 $\pm$ 10.79       &       -29.48 $\pm$ 11.23  \\
2459141.44400    &       -38.88 $\pm$ 10.55       &       -33.39 $\pm$ 11.35  \\
2459141.44920    &       -40.02 $\pm$ 10.58       &       -35.86 $\pm$ 10.39  \\
2459141.45685    &       -23.75 $\pm$ 12.46       &       -15.97 $\pm$ 11.48  \\
2459141.46185    &       -29.52 $\pm$ 11.09       &       -21.37 $\pm$ 10.61  \\
2459141.46801    &       -33.96 $\pm$ 11.97       &       -30.09 $\pm$ 11.43  \\
2459141.47317    &       -26.48 $\pm$ 10.66       &       -22.84 $\pm$ 11.26  \\
2459141.47954    &       -29.19 $\pm$ 11.41       &       -19.92 $\pm$ 13.34  \\
2459141.48453    &       -27.43 $\pm$ 12.16       &       -25.61 $\pm$ 10.99  \\
2459141.49077    &       -33.93 $\pm$ 11.18       &       -32.91 $\pm$ 10.22  \\
2459141.49586    &       -27.93 $\pm$ 13.34       &       -36.14 $\pm$ 12.77  \\
2459142.25492    &       ~19.34 $\pm$ 10.82       &       ~~2.72 $\pm$ 9.18   \\
2459142.26009    &       ~18.24 $\pm$ 10.84       &       ~~2.39 $\pm$ 8.65   \\
2459142.26518    &       ~~7.27 $\pm$ 10.22       &       ~-7.42 $\pm$ 8.81   \\
2459142.27040    &       ~22.76 $\pm$ 9.49        &       ~~3.06 $\pm$ 7.90   \\
2459142.27841    &       ~~9.91 $\pm$ 9.84        &       -14.59 $\pm$ 10.22  \\
2459142.28337    &       ~~3.55 $\pm$ 11.71       &       -11.81 $\pm$ 10.83  \\
2459142.28829    &       ~12.84 $\pm$ 8.58        &       ~-1.98 $\pm$ 9.59   \\
2459142.29390    &       ~11.85 $\pm$ 11.16       &       ~-6.01 $\pm$ 9.07   \\
2459142.29980    &       ~~7.30 $\pm$ 10.20       &       ~-5.24 $\pm$ 9.22   \\
2459142.30499    &       ~-3.88 $\pm$ 10.28       &       -17.11 $\pm$ 9.04   \\
2459142.31003    &       ~-5.55 $\pm$ 9.38        &       -14.60 $\pm$ 9.45   \\
2459142.31533    &       ~~5.89 $\pm$ 11.91       &       -13.03 $\pm$ 10.55  \\
2459142.32160    &       ~-7.84 $\pm$ 11.65       &       -13.08 $\pm$ 10.13  \\
2459142.32665    &       -11.09 $\pm$ 11.95       &       -13.09 $\pm$ 12.30  \\
\hline                                   
\end{tabular}
\end{table}

\newpage
\section{Model fit to all data}
Here we present the fitted RV+RM model, as well as the posterior distributions, where all the data points are considered. Namely, the four out of transit data points have not been masked.

\begin{figure}
    \centering
    \includegraphics[width=\linewidth]{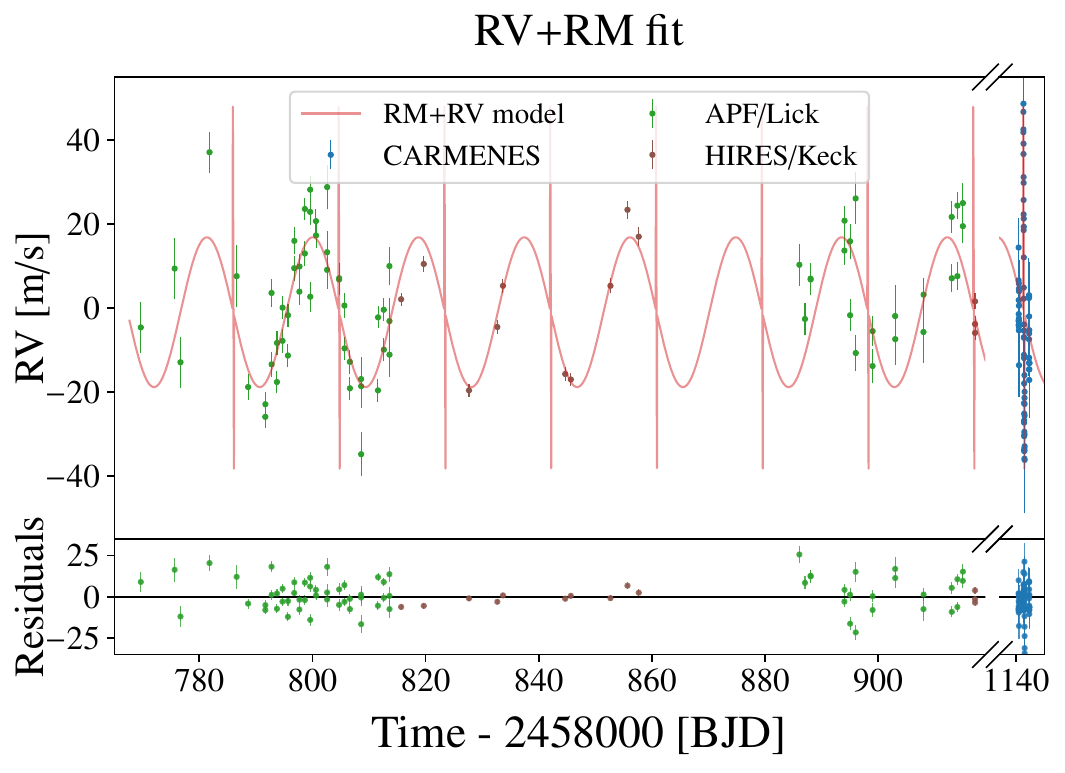}
    \includegraphics[width=\linewidth]{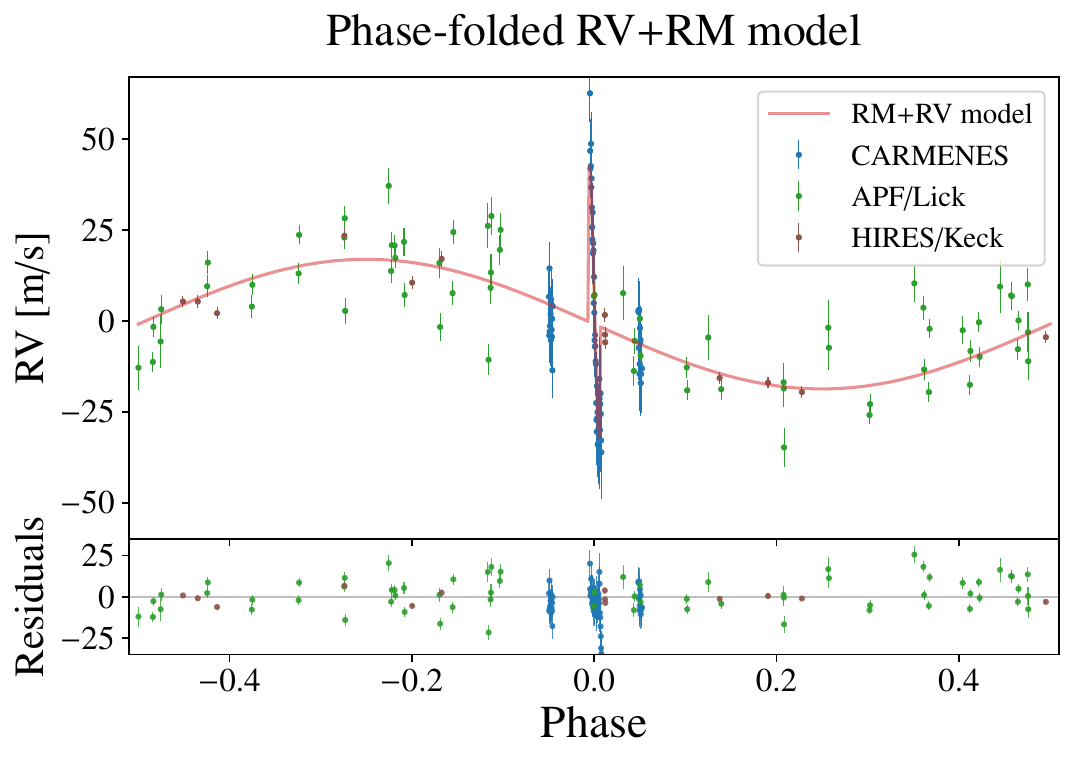}
    \includegraphics[width=\linewidth]{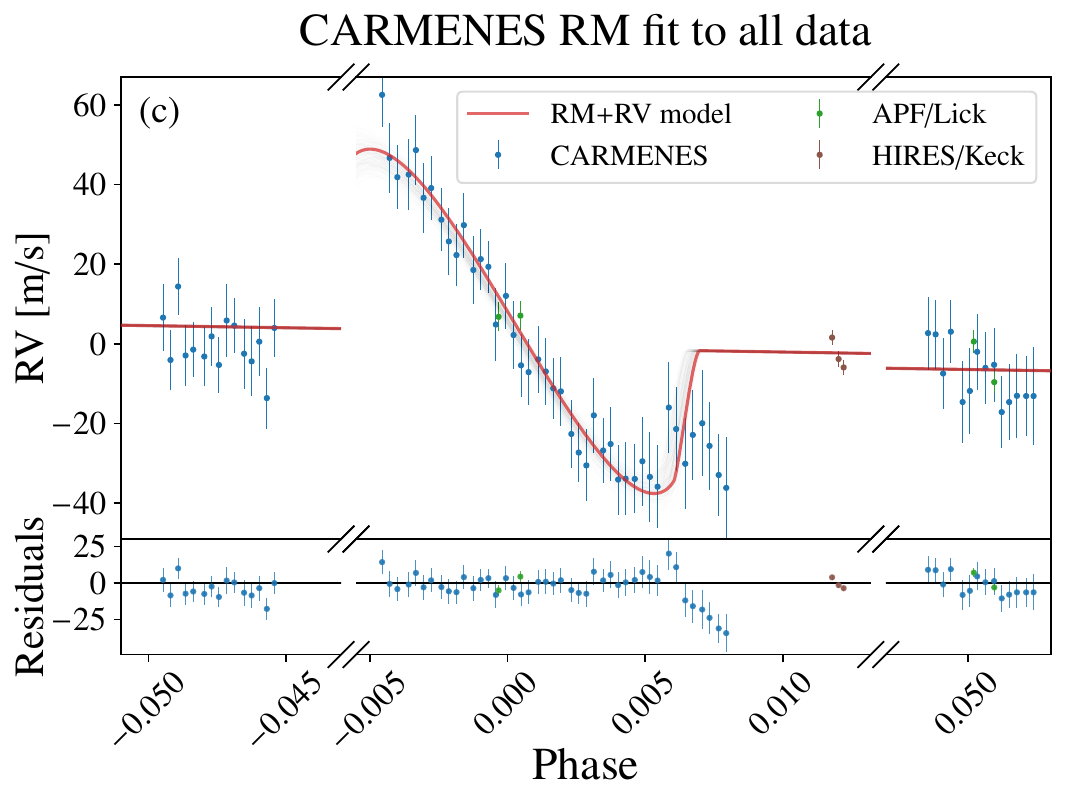}
    \caption{Same as Fig. \ref{fig: RV fit}, but with all the data points considered.}
    \label{fig: RV fit appx}
\end{figure}

\begin{figure}
    \centering
    \includegraphics[width=\linewidth]{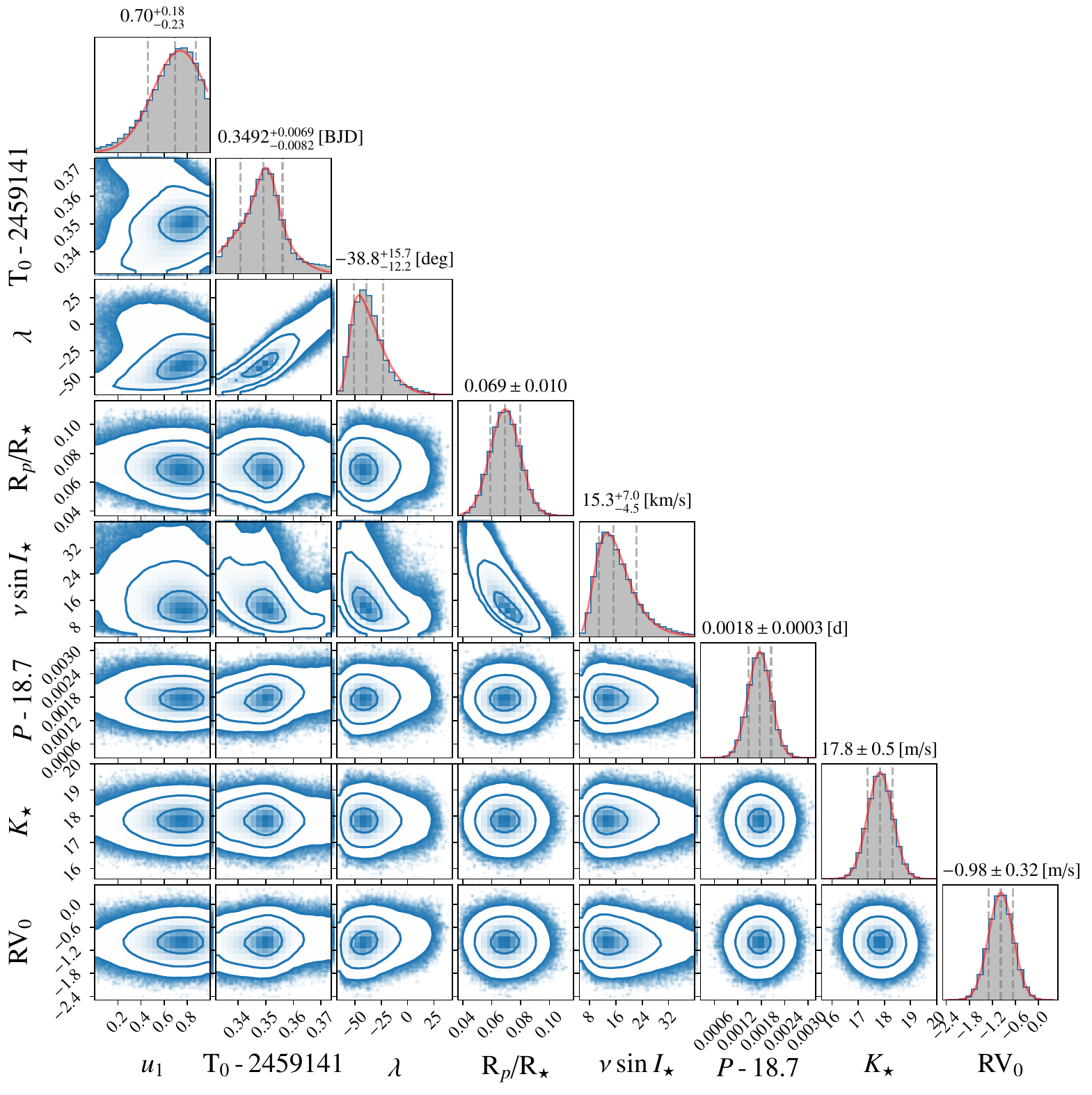}
    \caption{Same as Fig. \ref{fig: posteriors}, but for the fitting routine where all the data points are considered.}
    \label{fig: posteriors appx}
\end{figure}

\newpage
\section{Atmospheric analysis}
\subsection{Narrowband detections}
In Table \ref{tab: Narrowband detections} we summarise the list of species detected in exoplanetary atmospheres through the method of narrowband transmission spectroscopy, using high resolution échelle spectroscopy.
\renewcommand{\arraystretch}{1}
\begin{table*}[]
    \renewcommand\thetable{C.1}
    \centering
    \caption{Summary of detected species in exoplanetary atmospheres from the narrowband transmission spectroscopy method.}
    \begin{tabular}{l l l}
    \hline \hline
    Species & planet & Reference \\
    \hline
    Na\,{\scriptsize I} D & KELT-9\,b & \citet{Hoeijmakers2019}\\
    & KELT-20\,b & \citet{Casasayas2018,Casasayas2019,Hoeijmakers2020}\\
    & WASP-12\,b & \citet{Jensen2018}\\
    & WASP-49\,b & \citet{Wyttenbach2017}\\
    & WASP-69\,b & \citet{Casasayas2017}\\
    & WASP-76\,b & \citet{Seidel2019,Tabernero2021}\\
    & WASP-121\,b & \citet{Cabot2020,Hoeijmakers2020b,Borsa2021}\\
    & WASP-166\,b & \citet{Seidel2020}\\[4pt]
    K\,{\scriptsize I} \textlambda 7701 \AA & WASP-76\,b & \citet{Tabernero2021} \\
    & WASP-121\,b & \citet{Borsa2021}\\[4pt]
    Multiple Balmer lines & KELTP-9\,b & \citet{Wyttenbach2020}\\
    & KELT-20\,b & \citet{Casasayas2018,Casasayas2019}\\
    & WASP-33\,b & \citet{Borsa2021,Yan2021}\\
    H\textalpha & WASP-12\,b & \citet{Jensen2018}\\[4pt]
    He\,{\scriptsize I} IRT & GJ\,3470\,b & \citet{Palle2020}\\
    & HAT-P11\,b & \citet{Allart2018}\\
    & HD\,189733\,b & \citet{Salz2018}\\
    & HD\,209458\,b & \citet{Alonso2019}\\
    & WASP-69\,b & \citet{Nortmann2018}\\
    & WASP-107\,b & \citet{Allart2019,Kirk2020,Spake2021}\\
    \hline
    \multicolumn{3}{l}{This table has been compiled using the ExoAtmospheres database maintained by the Exoplanets and}\\
    \multicolumn{3}{l}{Astrobiology Group at the Instituto de Astrofísica de Canarias, which can be accessed at the following address:}\\
    \multicolumn{3}{l}{\url{http://research.iac.es/proyecto/exoatmospheres/index.php}}
    \end{tabular}
    \label{tab: Narrowband detections}
\end{table*}

\subsection{{\tt FastChem} abundances}
Here we present the equilibrium chemistry model calculated using the {\tt FastChem} code, for an atmosphere with solar C/O and metallicity, and $T_{eq}$ equal to that estimated for HD\,332231\,b, 876\,K.

\begin{figure*}
    \centering
    \includegraphics[width=\linewidth]{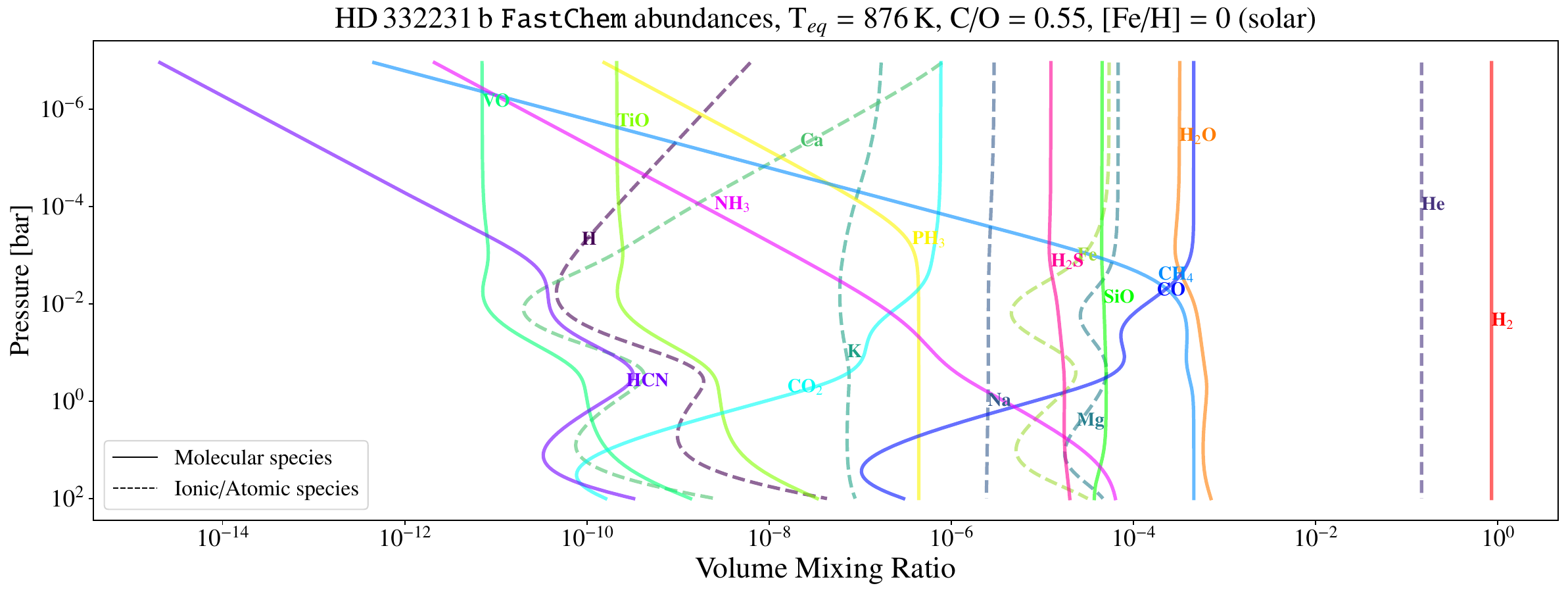}
    \caption{Equilibrium chemistry model for HD\,332231\,b calculated with {\tt FastChem}. The calculations are made for an atmosphere with $T_{eq}$ of 876\,K, and solar C/O and metallicity. For clarity the most abundant species are included, with the molecular species plotted as solid lines and those ionic and atomic species plotted with dashed lines.}
    \label{fig: FastChem abundances}
\end{figure*}

\end{appendix}
\end{document}